# SANDI: a compartment-based model for non-invasive apparent soma and neurite imaging by diffusion MRI

*Marco Palombo[1]\*, Andrada Ianus[1,2], Michele Guerreri[1], Daniel Nunes[2], Daniel C. Alexander[1], Noam Shemesh[2], Hui Zhang[1]*

[1] Centre for Medical Image Computing and Dept of Computer Science, University College London, London, UK.
[2] Champalimaud Research, Champalimaud Centre for the Unknown, Lisbon, Portugal.

*Abstract*

This work introduces a compartment-based model for apparent cell body (namely soma) and neurite density imaging (SANDI) using non-invasive diffusion-weighted MRI (DW-MRI). The existing conjecture in brain microstructure imaging through DW-MRI presents water diffusion in white (WM) and grey (GM) matter as restricted diffusion in neurites, modelled by infinite cylinders of null radius embedded in the hindered extra-neurite water. The extra-neurite pool in WM corresponds to water in the extra-axonal space, but in GM it combines water in the extra-cellular space with water in soma. While several studies showed that this microstructure model successfully describe DW-MRI data in WM and GM at b≤3,000 s/mm$^2$ (or 3 ms/µm$^2$), it has been also shown to fail in GM at high b values (b>>3,000 s/mm$^2$ or 3 ms/µm$^2$). Here we hypothesize that the unmodelled soma compartment (i.e. cell body of any brain cell type: from neuroglia to neurons) may be responsible for this failure and propose SANDI as a new model of brain microstructure where soma of any brain cell type is explicitly included. We assess the effects of size and density of soma on the direction-averaged DW-MRI signal at high b values and the regime of validity of the model using numerical simulations and comparison with experimental data from mouse ($b_{max}$ = 40,000 s/mm$^2$, or 40 ms/µm$^2$) and human ($b_{max}$ = 10,000 s/mm$^2$, or 10 ms/µm$^2$) brain. We show that SANDI defines new contrasts representing complementary information on the brain cyto- and myelo-architecture. Indeed, we show maps from 25 healthy human subjects of MR soma and neurite signal fractions, that remarkably mirror contrasts of histological images of brain cyto- and myelo-architecture. Although still under validation, SANDI might provide new insight into tissue architecture by introducing a new set of biomarkers of potential great value for biomedical applications and pure neuroscience.

\*Corresponding author: Dr. Marco Palombo, email: marco.palombo@ucl.ac.uk.

*1. Introduction*

Mapping brain microstructure noninvasively using diffusion-weighted MRI (DW-MRI) remains a formidable challenge due to the complexity of the underlying constituents and the relatively featureless diffusion-driven signal decay. Biophysical modelling can deliver more insight into the microstructure, thereby providing promising means for accessing MR-measurable parameters related to more specific features underpinning tissue or cellular structures. The existing conjecture or "standard model" of brain microstructure typically considers neural tissue as consisting of two compartments where endogenous water molecules diffuse (2-9): 1) a pool of water in neurites (axons, dendrites and neuroglial processes) thought to exhibit restricted diffusion and modelled by impermeable straight cylinders, or "sticks" if cylinder radius is assumed to be negligible (9); 2) another pool surrounding the neurites assumed to exhibit hindered diffusion and modelled as isotropic or anisotropic Gaussian diffusion (**Figure 1a**). The extra-neurite pool in white matter (WM) corresponds to water in the extra-axonal space, but in gray matter (GM) it combines water in the extra-cellular space with water in cell bodies of any brain cell type: from neuroglia to neurons (collectively named soma).

This work introduces a biophysical model incorporating for the first time soma size and density in addition to neurite density, thereby enabling their joint estimation non-invasively using DW-MRI and a model-based approach. The model is motivated by recent studies that suggest the standard model of neural tissue microstructure (2-8) does not hold in GM at high b-values (10-15). We hypothesise that the observed departure of the standard model from the data at high b-values in GM can be largely explained by the breakdown of the assumption that water in soma exhibits similar diffusion properties as water in extra-cellular space (12, 16). Furthermore, we propose that this unexplained signal can be accounted for by explicitly modelling the soma as one of the contributors to the intra-cellular signal (**Figure 1b**). Most importantly, the resulting biophysical model enables us to estimate apparent soma size and density non-invasively using DW-MRI.

We tested our hypothesis using Monte-Carlo diffusion simulations in simplified digital models of neural cells (1). We show that soma size and density have indeed a specific signature on the direction-averaged (also known as powder-averaged (17)) DW-MRI signal at high b-values that are consistent with the observed departure. Furthermore, using the same Monte-Carlo simulation framework, we show for the first time that, at reasonably short diffusion times ($t_d$) of few tens of milliseconds, the water exchange between neurites and soma can be ignored, supporting the design of a simple three-compartment model to separate and quantify the presence of soma (**Figure 1b**). We note that it is still not clear whether using DW-MRI we can distinguish neuroglial from neuronal signal; therefore, we expect our model likely quantifies the presence of cell bodies of all cell types in the brain (e.g. neuroglia and neurons). We evaluate the resulting model with data from healthy ex-vivo mouse brain and in-vivo human brains, with results supporting the model as a promising tool for estimating apparent soma size and density.

The rest of the paper is organised as follows: in the Theory section we briefly recall the current standard model of brain microstructure and we formally introduce a new biophysical model of brain microstructure, explicitly accounting for the soma compartment and corresponding signal. In the Methods section we describe the numerical simulations and experiments used in this work to support such model, and in the Results section, we show promising results on how the proposed model enables us to characterize both cyto and myeloarchitectonic of the brain non-invasively using DW-MRI. We finally discuss the results, as well as the model limitations in the Discussion section.

## 2. *Theory*

The current paradigm in model-based microstructure imaging uses biophysical models, inspired by microscopy studies of tissue microarchitecture, to approximate the tissue microenvironment and estimate model parameters linked to specific tissue microstructure features from DW-MRI data (6-8). Among them, the most common class of models separate the contribution to the DW-MRI signal *S* measured in an image voxel into different parts that can be attributed to different "compartments" where water molecules diffuse (9). A common assumption of these so-called compartment models of diffusion is that there is no exchange between the compartments, i.e., water molecules do not move from one compartment to the other.

In the simplest form, two compartments are typically used and are identified as *intra-* and *extra-cellular*. For neural tissue, it is common to assume the signal associated with the intra-cellular compartment ($S_{intra}$) is mostly due to water diffusing in elongated cellular fibres, generally called neurites (2-8), modelled as straight cylinders of zero diameter, namely sticks. Such models merge any signal contribution from soma or other large cellular domains with that of the extra-cellular compartment ($S_{extra}$). Below, the detail of this standard model is first given, before we describe the proposed extension that models soma explicitly.

*2.1 The Standard Model of Neural Tissue Microstructure*

Following (3, 18), we briefly recall here that in general the normalised DW-MRI signal associated with the intra-neurite compartment can be represented as a convolution between the fibre orientation distribution function (fODF) $P(\hat{\boldsymbol{n}})$ and the response kernel K from a perfectly aligned fiber (fascicle) pointing in the direction $\hat{\boldsymbol{n}}$, such that (3, 18)

$$A_{in}(\boldsymbol{b}) = \int_{|\hat{\boldsymbol{n}}|=1} d\hat{\boldsymbol{n}}\, P(\hat{\boldsymbol{n}})\, K(b, \hat{\boldsymbol{n}} \cdot \hat{\boldsymbol{g}}) \quad (1)$$

where b is the diffusion weighting factor, measured along the direction $\hat{\boldsymbol{g}}$, such that $\boldsymbol{b} = b\hat{\boldsymbol{g}}$. The "stick" model assumes the functional form for the response kernel:

$$K(b, \hat{\boldsymbol{n}} \cdot \hat{\boldsymbol{g}}) = e^{-bD_{in}^{\perp} - b\left(D_{in}^{\parallel} - D_{in}^{\perp}\right)(\hat{\boldsymbol{n}} \cdot \hat{\boldsymbol{g}})^2} \quad (2)$$

modelled by axially symmetric Gaussian diffusion compartment, with the radial diffusion of the intra-neurite compartment $D_{intra}^{\perp} = 0$ (i.e. sticks). Starting from this common paradigm for the intra-neurite signal, many methods have been proposed to quantify neurite density and dispersion in both WM and GM (2-8, 18-26). They mostly differ in the way they model the fODF or in the way they model the contribution to the total signal from the extra-neurite compartment. Nevertheless, they all can be seen as multi-compartment models (**Figure 1a**) where the total signal of the tissue component measured in an imaging voxel is given by:

$$\frac{S(\boldsymbol{b})}{S(0)} = f_{in} A_{in}(\boldsymbol{b}) + (1 - f_{in}) A_{en}(\boldsymbol{b}) \quad (3)$$

The signal from extra-neurite compartment $A_{en}(\boldsymbol{b})$ is modelled as an isotropic or anisotropic diffusion tensor, with its principal direction of diffusion assumed to be parallel with the dominant direction of the fODF. Since intra- and extra-neurite compartments may generally have different $T_2$ values, the fraction $f_{in}$ is the relative signal fraction, not the absolute volume fraction (27). Moreover, the myelin water contribution is assumed unobservable due to its short $T_2$ time compared to clinical DW-MRI echo time TE (28). Also, further compartments, such as isotropic cerebrospinal fluid (CSF), can be added to Eq. (3) to accommodate partial volume contamination, such as in (3).

## 2.2 SANDI: Microstructure Model for Soma Imaging

Here, we propose an intra-cellular model that consists of the intra-neurite model accommodating an approximate description of the contribution from water spins diffusing within cellular soma (**Figure 1b**).

**Model assumptions.** The proposed microstructural model is based on the same assumptions of the "standard" model and on the experimental evidence that at short $t_d$ ($\leq$20 ms given a water bulk diffusivity of ~3 $\mu m^2$/ms and estimated pre-exchange time $\geq$500 ms) the effect of cell (either neurons or glia) membrane permeability and corresponding water exchange between intra- and extra-cellular space is negligible (29). An additional assumption, whose validity is investigated in this work by numerical simulations, is that at short $t_d$ ($\leq$20 ms), the two sub-compartments comprising the intra-cellular space: soma and neurites, can be approximated as two non-exchanging compartments.

**General formulation.** Under these assumptions, we propose the functional form for the new compartment model of brain tissue microstructure to be

$$\frac{S(\boldsymbol{b})}{S(0)} = f_{ic}(f_{in}A_{in}(\boldsymbol{b}) + f_{is}A_{is}(\boldsymbol{b})) + f_{ec}A_{ec}(\boldsymbol{b}) \quad (4)$$

where $f_{ic}$ and $f_{ec}$ are the intra-cellular and extra-cellular relative signal fractions satisfying the condition $f_{ic} + f_{ec} = 1$; $f_{in}$ and $f_{is}$ are the neurite and soma relative signal fractions satisfying the condition $f_{in} + f_{is} = 1$; $A_{in}$ and $A_{is}$ are the normalised signals for restricted diffusion within neurites and soma, respectively and $A_{ec}$ the normalised signal of the extra-cellular space. Equation (4) represents a first attempt to provide a more general model to describe neural tissue microstructure that takes into account soma. Indeed, following a hierarchical decomposition of the tissue compartments similar to previous works (2-5, 23, 30-33), we identify an intra-cellular and an extra-cellular compartment, contributing to the total signal with relative signal fractions $f_{ic}$ and $f_{ec}$ that have to sum up to unity (**Figure 1b**). Then, the intra-cellular compartment is comprised of intra-neurite and intra-soma compartment that contribute to the total intra-cellular signal with relative signal fractions $f_{in}$ and $f_{is}$ that have to sum up to unity (**Figure 1b**). As such, $f_{ic}$, $f_{ec}$, $f_{in}$ and $f_{is}$ are not volume fractions of the corresponding constituent of the MRI voxel, but rather relative MRI signal fractions of the corresponding tissue compartment. In fact, the T2 of the intra- and extra-cellular compartments may be different (27, 34), and here we are additionally neglecting myelin water assuming the echo time is sufficiently long to attenuate most of the myelin water contribution through relaxation. Moreover, in this first implementation, CSF contributions are not taken into account since; however, due to its quickly decaying signal with increasing b values we expect that its residual contribution would be simply captured by the extra-cellular compartment and would not significantly impact the estimates of the intra-cellular compartment model parameters.

**Direction-average.** Here, we focus on estimating orientation-independent features of microstructure by considering the direction-averaged signal $\tilde{S}(b)$ (35). The direction-averaged signal, also known as the powder averaged, is defined as the average of the signals $S(\boldsymbol{b})$ acquired along many uniformly distributed directions $\hat{\boldsymbol{g}}$. The resulting signal takes the form:

$$\frac{\tilde{S}(b)}{S(0)} = (1 - f_{ec})(f_{in}\tilde{A}_{in}(b) + (1 - f_{in})\tilde{A}_{is}(b)) + f_{ec}\tilde{A}_{ec}(b) \quad (5)$$

where $\tilde{A}_{in}$, $\tilde{A}_{is}$, and $\tilde{A}_{ec}$ are the direction-averaged normalised signals associated with their respective compartments and we used the relations $f_{ic}$ = 1 - $f_{ec}$ and $f_{is}$ = 1 - $f_{in}$. The direction averaging eliminates the dependence on the fODF which is readily determined following the estimation of the orientation-independent microstructure features (5, 35-37). The forms of these direction-averaged signals are given below.

*Extra-cellular compartment.* The diffusion of water molecules associated with the extra-cellular compartment, $\tilde{A}_{ec}$, follows the assumptions made under the standard model. It is modelled as isotropic Gaussian diffusion with a scalar effective diffusion constant $D_{ec}$:

$$A_{ec}(b, D_{ec}) = \tilde{A}_{ec}(b, D_{ec}) \approx e^{-bD_{ec}} \quad (6)$$

This approximation assumes that the extra-cellular space is fully connected for water molecules to sample during the course of the diffusion experiment.

*Intra-neurite compartment.* The signal contribution $\tilde{A}_{in}$ from neurites (dendrites and axons) also follows the standard model. On a voxel level, it is assumed that neuronal processes can be described as a collection of long thin cylinders, with a longitudinal apparent diffusion coefficient $D_{in}^{\parallel} \equiv D_{in}$ and a transverse apparent diffusion coefficient $D_{in}^{\perp} \sim 0$. It is also assumed that on the timescale of our diffusion experiments ($t_d$ ~ 10 ms), the effects of branching and/or finite length of cellular processes can be neglected. These assumptions are appropriate under the considered experimental conditions. The root-mean-squared-displacement along the neurite would be ~5 μm, for typical longitudinal intra-neurite diffusivity which is half that of free water at body temperature. This distance is much smaller than the typical length of each cell fibre's branch, e.g. ~55 μm for cerebral cortical pyramidal neurons (38, 39). Therefore, from the water diffusion standpoint, we can consider branching neurites as a collection of individual branches (or cylinders/sticks), randomly oriented in space (38, 40). Under these assumptions, the direction-averaged $A_{in}$ can be computed as powder average of randomly oriented sticks, such that (2, 5, 9, 21, 35-37):

$$\tilde{A}_{in}(b, D_{in}) \approx \sqrt{\frac{\pi}{4bD_{in}}} erf(\sqrt{bD_{in}}) \qquad (7)$$

*Intra-soma compartment.* The signal contribution $\tilde{A}_{is}$ from cell bodies is assumed to arise from a pool of diffusing water molecules restricted in spheres of radius $r_s$. With this approximation, we are implicitly assuming negligible exchange between the pool of diffusing water molecules confined in the intra-neurite space and those in the soma. While this approximation is not valid in general, we will show, using Monte Carlo diffusion simulation in realistic models of neuronal cells, that under practical experimental conditions and typical soma size and volume fraction, the exchange between intra-neurite and intra-soma diffusing water is negligible.

Modelling soma as closed impermeable spheres, the normalised signal can be computed from the GPD approximation (41, 42), such that

$$\tilde{A}_{is}(b, D_{is}, r_s) \approx exp\left\{-\frac{2(\gamma g)^2}{D_{is}} \sum_{m=1}^{\infty} \frac{\alpha_m^{-4}}{\alpha_m^2 r_s^2 - 2}\right.$$
$$\times \left[2\delta - \frac{2 + e^{-\alpha_m^2 D_{is}(\Delta-\delta)} - 2e^{-\alpha_m^2 D_{is}\delta} - 2e^{-\alpha_m^2 D_{is}\Delta} + e^{-\alpha_m^2 D_{is}(\Delta+\delta)}}{\alpha_m^2 D_{is}}\right]\right\} \qquad (8)$$

where $D_{is}$ is the bulk diffusivity of water in somas, δ and Δ the diffusion gradient pulse width and separation, g the magnitude of diffusion gradient pulse, $\alpha_m$ the m$^{th}$ root of the equation $(\alpha r_s)^{-1} J_{\frac{3}{2}}(\alpha r_s) = J_{\frac{5}{2}}(\alpha r_s)$, with $J_n(x)$ the Bessel function of the first kind. For simplicity, here we consider a single radius $r_s$ as representative for all the soma in a given MRI voxel. In reality, we would expect a distribution of radii P($r_s$) in a given MRI voxel of few millimetres. In this case, the normalised signal could be computed from Eq. (8) following a volume average (43):

$$\tilde{A}_{is}(b, D_{is}) = \frac{\int_0^\infty P(r_s) r_s^3 \tilde{A}_{is}(b, D_{is}, r_s) dr_s}{\int_0^\infty P(r_s) r_s^3 dr_s} \qquad (9)$$

The $r_s^3$ term is included to account for the spin volume, i.e., the increase in the number of spins as the radius increases. In principle, it is possible to use Eq. (9) to analyze experimental data. However, the inversion of Eq. (9), a Fredholm equation of the first kind, to provide P(r$_s$) is non-trivial, and to obtain an approximate estimate of apparent soma size, we prefer to use in this work Eq. (8).

*Total signal.* Substituting Eqs. (6), (7) and (8) in Eq. (5), we get the approximated expression for the total direction-averaged signal:

$$\frac{\tilde{S}(b)}{S(0)} = (1 - f_{ec})(f_{in}\tilde{A}_{in}(b, D_{in}) + (1 - f_{in})\tilde{A}_{is}(b, D_{is}, r_s)) + f_{ec}\tilde{S}_{ec}(b, D_{ec}) \quad (10)$$

In general, the total free parameters to be determined from the direction-averaged data are thus six: $f_{in}$, $f_{ec}$, $D_{in}$, $D_{is}$, $D_{ec}$ and $r_s$. However, from Eqn. (8), it is evident that it is challenging to disentangle $D_{is}$ from $r_s$ in many practical applications where data are usually acquired by varying only the magnitude of diffusion gradient; thus, it is possible to estimate only the ratio $D_{is}/r_s$, or $D_{is}$ from fixing $r_s$ or $r_s$ from fixing $D_{is}$. Since the purpose of the proposed biophysical model is to characterize the microarchitecture of the brain tissue, an apparent MR estimate of $r_s$ would be more valuable than that of $D_{is}$. Therefore, we will adopt the simplification of fixing $D_{is}$ to the value of the self-diffusion coefficient of free water, given the tissue temperature. Since all the experiments in this study were conducted *in vivo,* or *ex vivo* with the temperature kept constant at 37 °C, for fitting purposes, we will fix $D_{is} \sim 3$ μm$^2$/ms. However, in typical diffusion experiments, fixing $D_{is}$ to any value between 0.5 and 3 μm$^2$/ms would not change the estimates of r$_s$ substantially, as suggested by previous studies using numerical simulations and PGSE experiments in murine erythroleukemia cancer cells (44).

### 3. Methods

#### 3.1 Numerical Simulations

Monte-Carlo simulation of spin-diffusion in realistic digital models of dendritic structures were conducted to investigate the regime of validity of the assumption of non-exchanging intra-cellular compartments. Since the purpose of the simulations is to investigate when the assumption of non-exchanging neurite and soma compartments holds, only the intra-cellular component of the total MR signal is of interest. We first establish the regime of validity of our model in Eq. (10) and then further use the simulations to investigate the sensitivity to soma size and density (r$_s$, f$_{is}$) within that regime.

*Simulation setup.* Detailed 3D geometries were constructed using our recently proposed generative model of complex cellular morphologies (1) that enables users to simulate molecular diffusion within realistic digital brain cells, such as neurons, in a completely controlled and flexible fashion (**Figure 2a** and **b**). Here we use the generative model to mimic realistically connected neurites with different ($r_s$, $f_{is}$) combinations. We assume cell fibres do not branch and simulate only the intra-cellular signal (**Figure 2c**) – hence, the generative model in **Figure 2a** are not tested in simulation. Therefore, the experiments test the validity of only the assumptions of the intra-cellular compartment models: intra-neurites as randomly oriented sticks, intra-soma as sphere, and negligible exchange of diffusing spins between them. Specifically, we used 20 randomly oriented straight cylindrical segments of radius $r_n$ = 0.50 µm and length L = [50, 200, 500] µm, and spherical soma of radius $r_s$ = [2, 4, 6, 8, 10] µm leading to a fraction of the total cell volume occupied by the soma, $f_{is} = \frac{4/3\pi r_s^3}{20 \times L \times \pi r_n^2}$ ranging from ~ 0.1-0.9, to simulate structures mimicking a range of possible brain cell types, from small microglia to large neurons. We note that radius of brain neurites (glial processes, dendrites, axons and neuropil in general) is typically ≤ 1.5 µm (45). For such very thin fibres, we do not expect significant effects on the DW-MRI signal measured by sequences like Pulsed Gradient Spin Echo (PGSE) at the experimental conditions investigated here (46). The diffusion of $5 \times 10^5$ non-interacting spins, initially uniformly distributed within the whole cellular volume, was simulated for each synthetic geometry with bulk diffusivity 2 µm²/ms and time step 20 µs, using CAMINO (47). The number of spins and the time step were chosen as the minimal values that guarantee stability of the simulated signal, according to previous work (48). Using more spins or smaller time step would produce an identical simulated signal within <2% of error. To investigate the validity of the non-exchanging intra-cellular compartments approximation, a set of 3D digital models were created to explicitly prevent any exchange between soma and neurites. This is done by sealing off all the holes in the surface mesh of each sphere used to model soma (**Figure 2c**), thus disconnecting each sphere from the neurites that extend from it, which are modelled by cylinders. Note that in these simulations we used simplified brain cell structures with non-branching neurites. This is adequate, because, as justified earlier, branching neurites can be approximated as a collection of individual non-exchanging branches under the considered experimental conditions.

*Simulation 1 – Investigating the validity of the non-exchange approximation for different diffusion times.* The purpose of this simulation experiment is to theoretically investigate using simulations when, in terms of chosen diffusion time, the non-exchanging neurite/soma compartments assumption used to build SANDI model holds. From the simulated spin-trajectories, the normalized direction-averaged DW-MRI signal $\tilde{A}$ was computed from a PGSE sequence with $t_d$ ranging from 1 to 240 ms, δ = 1 ms and three b values: 500, 1,000 and 2,000 s/mm² (or, 0.5, 1 and 2 ms/µm²). This resulted in two sets of normalized direction-averaged signals: with exchange ($\tilde{A}_w$) and without exchange ($\tilde{A}_{w/o}$) that were used to compute the corresponding apparent diffusion coefficients $ADC_w$ and $ADC_{w/o}$ for different $t_d$. The relative difference ($\Delta ADC$) between $ADC_w$ and $ADC_{w/o}$ was computed as a function of $t_d$ according to the following definition:

$$\Delta ADC(t_d) = \left|\frac{ADC_{w/o}(t_d) - ADC_w(t_d)}{ADC_{w/o}(t_d)}\right| \times 100 \qquad (11)$$

A sensible regime where the non-exchanging-compartments approximation can be considered valid may be for those values of $t_d$ where $\Delta ADC(t_d) < 10\%$.

*Simulation 2 – Investigating the validity of the non-exchange approximation for different b values.* The purpose of this simulation experiment is to theoretically investigate using simulations whether the non-exchanging neurite/soma compartments assumption holds for a wide range of (high) b values, when $t_d$ is fixed to a short or long value. From the simulated spins' trajectories, the normalized direction-averaged DW-MRI signal $\tilde{A}$ was computed from a PGSE sequence with b-values = [0 : 1,000 : 60,000] s/mm² (or [0:1:60] ms/μm²) and 32 directions, uniformly distributed over the full sphere. Gradient pulse duration $\delta$ = 3 ms and separation $\Delta$ = 11 and 81 ms, were chosen according to the results of Simulation 1 (see Results section) and to match experimental data (see following section 3.2). Furthermore, in order to quantify the bias in model-parameter estimation due to the non-exchanging assumption, Eq. (10) without the extra-cellular compartment was fitted to the simulated signals with exchange between neurites and soma. The uncertainty in parameter estimation was evaluated with a Monte Carlo approach. Specifically, the residual sum of squares corresponding to the best initial fit for each ($r_s$, $f_{is}$) configurations was used as standard deviation to randomly induce artificial Gaussian noise in our simulated signals before repeating the fitting operation. This process was performed 1,000 times. Then for each parameter, we computed its mean and standard deviation over the generated repetitions, and compared them with the ground-truth values. We performed this analysis for both conditions: $t_d$ = 10 ms (when the non-exchange assumption should hold, according to Simulation 1) and $t_d$ = 80 ms (when the non-exchange assumption should fail, according to Simulation 1).

*Simulation 3 – Investigating the sensitivity of signal to soma size and density.* The purpose of this simulation experiment is to investigate using simulation and comparison with real data the sensitivity of signal to soma size and density. Using the result in Simulation 2, we built a dictionary of simulated signals, corresponding to different microstructural scenarios, i.e. different ($r_s$, $f_{is}$) configurations, and compared them to the experimental signals obtained from selected regions-of-interest (ROIs) in ex-vivo mouse brain (see following section 3.2). The direction-averaged DW-MRI signal from real data was computed for a GM ROI manually drawn in the cortex and a WM ROI in the corpus callosum. We chose these two brain regions because they are expected to have very different ($r_s$, $f_{is}$) values. The average signal in each ROI was compared against the dictionary of simulated signals to determine whether different soma size and density could explain the non-exponential signal decay in experimental data.

### *3.2 Experimental Data*

We considered two datasets to evaluate our key modelling assumption – diffusion of water molecules within soma has a non-negligible contribution to the normalised direction-averaged signal at high b values and can be modelled as restricted diffusion, separate from water diffusion in neurites – and to assess the proposed model's ability to estimate soma size and density. First, DW-MRI data of ex-vivo mouse brains were collected at ultra-high b values, with state-of-the-art preclinical hardware, to show that the specific signature of soma size and density on the DW-MRI signal, as predicted by Monte Carlo simulation (Simulation 3), is consistent with measured data. Second, DW-MRI data of in-vivo healthy human brains at high b values were analysed, producing maps of soma density that can be compared against published histological results, to show that the technique translates to in-vivo human studies and that it provides a novel contrast sensitive to neural tissue cytoarchitecture.

**Ex-vivo mouse brain.** All animal experiments were preapproved by the institutional and national authorities and were carried out according to European Directive 2010/63. A c57bl/6 mouse (N=1), male, 8 weeks old, was perfused intracardially with 4% paraformaldehyde. The brain was isolated and kept 48h in 4% paraformaldehyde and 5 days in PBS (changed daily), before being transferred to a 10 mm NMR tube filled with Fluorinert (Sigma Aldrich) for susceptibility matching. MRI experiments were performed using a 16.4 T MRI scanner (Bruker BioSpin, Karlsruhe, Germany) operating at 700 MHz for $^1$H nuclei and equipped with a micro5 imaging probe (Bruker BioSpin, Rheinstetten, Germany) with maximum gradient strength 3000 mT/m isotropically. The brain was kept at constant temperature of 37°C using the probe's temperature controller. DW-MRI were acquired using a PGSE-EPI sequence with: TE/TR=20/2500 ms; $\delta/\Delta$=3/11 ms; 30 b values from 1 to 40 ms/$\mu$m$^2$; 40 gradient directions per b value, 30 b=0 images, slice thickness = 0.250 mm; FOV = 11.2x11.2 mm; matrix dimension = 224x224; bandwidth ~ 250 kHz; resolution 50x50x250 $\mu$m$^3$, 10 slices, 4 averages. The dataset was denoised using MRtrix3 (49) (http://www.mrtrix.org) and corrected for Gibbs ringing (50). No artifacts from eddy-current were observed. The direction-averaged DW-MRI signal was then computed for a GM ROI manually drawn in the cortex and a WM ROI in the corpus callosum. The average signal in each ROI was compared against the dictionary of simulated signals in Simulation 3 to determine whether different soma size and density could explain the non-exponential signal decay in experimental data.

**In-vivo human brain**. To provide proof-of-concept of translation to in-vivo human applications, we performed a retrospective analysis of 25 healthy young subjects (age between 25 and 35) from the MGH Adult Diffusion Dataset downloaded from the HCP data repository (https://www.humanconnectome.org). While this dataset was not acquired with the present application in mind, its sequence parameters turn out to be almost optimal for sensitivity of the direction-averaged signal to soma according to our model, i.e. $t_d$<20 ms and many b-values >3 ms/$\mu m^2$. The dataset was acquired on a 3T Siemens Connectom scanner, customized with a 64 channel tight-fitting brain array coil (51) and consists of MPRAGE and diffusion scans with four levels of diffusion weighting. The b-values used were 1, 3, 5 and two acquisitions at 10 ms/$\mu m^2$ with respectively 64, 64, 128, 128 and 128 randomly distributed diffusion-encoding directions over a full sphere. The signal-to-noise ratio (SNR) of individual b = 10 ms/$\mu m^2$ images was ~ 5 and the SNR of the direction-averaged images at b = 10 ms/$\mu m^2$ was ~ 50. Every 14th volume was an image without diffusion weighting (b0) used for motion correction and normalisation. Other acquisition parameters were TE = 57 ms, TR = 8800 ms, $\delta$ = 13 ms, $\Delta$ = 22 ms, voxel size = 1.5 mm$^3$ isotropic, field of view = 210 × 210 mm$^2$, pixel bandwidth = 1984 Hz/Px, echo spacing = 0.63 ms and parallel imaging factor = 3. Additional scan details can be found in (52). The DW-MRI data in the dataset were already pre-processed with software tools in FreeSurfer V5.3.0 (http://freesurfer.net/fswiki/FreeSurferWiki/) and FSL V5.0 (http://fsl.fmrib.ox.ac.uk/fsl/fslwiki/). Specifically, the distortion caused by the gradient nonlinearity was corrected based on the spherical harmonic coefficients (53). For motion correction, the b=0 images interspersed throughout the diffusion scans were used to estimate the bulk head motions with respect to the initial time point (first b=0 image), where the rigid transformation were calculated with the boundary based registration tool in the FreeSurfer package V5.3.0 (54). For each b=0 image, this transformation was then applied to itself and the following 13 diffusion weighted images to correct for head motions. The FSL's 'EDDY' tool was to correct for eddy current distortion. All 4 shells (b = 1, 3, 5, 10 ms/$\mu m^2$) were concatenated (552 volumes in total) and passed into the EDDY tool. After eddy current correction, the rigid rotational motion estimates obtained from both the motion correction step and the eddy current correction step were concatenated and applied to the original b-vectors for correction.

### 3.3 Model-parameter estimation

The five model parameters: $f_{in}$, $f_{ec}$, $D_{in}$, $D_{ec}$ and $r_s$ are estimated by random forest regression (55, 56), while $f_{is}$ is computed from the relation $f_{is} = 1 - f_{in}$. To train the random forest regressor, $10^5$ synthetic signals were generated using Eq. (10), with $10^5$ random values of the five model parameters chosen uniformly distributed within the reasonable intervals: $f_{in}$=[0.01, 0.99]; $f_{ec}$=[0.01, 0.99]; $D_{in}$=[0.1, 3] µm²/ms; $D_{ec}$=[0.1, 3] µm²/ms; $r_s$=[1, 12] µm. For testing, we used 2x10⁴ previously unseen signals generated in the similar way. To match the SNR of the signals to be fitted, Rician-distributed noise was added to the synthetic data used for training and testing. We implemented a random forest regressor using the scikit-learn open source Python toolkit (57). Following preliminary experiments, we built the final random forest regressor with 200 trees of maximum depth 20 and bagging as the setting that maximises the performance of the model. Further general implementation details can be found at http://scikit-learn.org/.

*Accuracy and precision of intra-cellular model-parameter estimation.* This section investigates the robustness of the machine-learning based fitting algorithm we use. We explored different ($r_s$, $f_{is}$) parameters combinations, using the simulated intra-cellular signals in Simulation 2, with $r_s$ = [2, 4, 6, 8, 10] µm and $f_{is}$ = [0.01 0.02 0.05 0.15 0.30 0.45 0.60 0.65 0.85]. For each combination we estimated the three free parameters $f_{in}$, $D_{in}$ and $r_s$ (since we are not considering the extra-cellular contribution) by RF regression, and repeated the experiment 2500 times with different noise instances to estimate mean and variance of the parameter estimation and thus quantify accuracy (through bias) and precision (trough statistical dispersion or standard deviation). The $f_{is}$ was estimated by the relation $f_{is} = 1 - f_{in}$. Different amount of Rician distributed noise was added to the simulated intra-cellular signals in Experiment 2 by adding complex Gaussian noise before computing the magnitude to simulate three SNR conditions: SNR = 10 (worse scenario), SNR = 50 (similar to our experimental SNR) and ∞ (i.e. no noise, ideal scenario).

Additionally, for the ideal scenario of SNR = ∞, we also performed an ablation study to assess to what extent the accuracy and precision are compromised by using less and/or lower b values than those in Simulation 2. We tested four different combinations of b values, subsampled from Simulation 2 at $t_d$=10 ms, that could be achieved by clinical scanners or more powerful human scanners such as the Connectom (58) : b = [0, 0.7, 1.5, 2, 3] ms/µm²; b = [0, 0.7, 1.5, 3, 10] ms/µm²; b = [0, 1, 3, 5, 10] ms/µm²; b = [0, 1, 2, 3, 5, 10, 25] ms/µm². We explored the same set of ($r_s$, $f_{is}$) parameter combinations as above. For each combination, we estimated the three free parameters $f_{in}$, $D_{in}$ and $r_s$ in the same way as described above, for each b combination. We computed the mean squared error (MSE) to quantify the overall changes in accuracy and precision for each estimate compared to the ground-truth values, known by design.

*3.4 Comparison with dot-compartment model*

Because it has been reported (9, 31, 32) that in fixed tissue a fraction of immobile water, known as the "dot-compartment", is not negligible for WM, we compared SANDI to a variant that replaces the sphere compartment with the simpler dot-compartment, to assess which one describes better the high b-value data, in both GM and WM. We fitted to the experimental data from ex-vivo mouse brain both SANDI (Eq. (10)) and its dot-compartment variant where the sphere compartment in Eq.(10) has been substituted by a dot-compartment of relative signal fraction $f_{dot}$ = 1- $f_{in}$; specifically:

$$\frac{\tilde{S}(b)}{S(0)} = (1 - f_{ec})(f_{in}\tilde{A}_{in}(b, D_{in}) + f_{dot}) + f_{ec}\tilde{S}_{ec}(b, D_{ec}) \quad (12)$$

The corrected Akaike's information criterion (AICc) (59) was used to compare the relative fit quality of the two models. Given a set of candidate models for the data, the preferred model is the one with the lowest AICc value. The models' degrees of freedom were 5 for SANDI (Eq.(10)) and 4 for the dot-compartment variant (Eq.(12)).

*3.5 Comparison with histology from literature*

To illustrate qualitatively that the contrasts in $f_{in}$ and $f_{is}$ maps mirror the myelo- and cyto-architecture of the brain, $f_{in}$ and $f_{is}$ maps for one representative subject were qualitatively compared against literature-derived histological images of myelin- and Nissl-stained sections of the human brain from https://msu.edu/~brains/brains/human.

Furthermore, parametric maps of $f_{is}$ for each of the 25 analysed subjects were processed with FreeSurfer Software Suite (https://surfer.nmr.mgh.harvard.edu) and projected onto the inflated cortical surface extracted from each corresponding subject for visualisation. Projection of the average $f_{is}$ map across all the 25 subjects onto a common template (cortical surface-based atlas defined in FreeSurfer based on average folding patterns mapped to a sphere) was also computed and the parcellation of the cortical surface according to Brodmann areas (BA) was performed with FreeSurfer. Brodmann parcellation was chosen because it is based on differences in brain cytoarchitecture features, making it ideal to show correspondence between the proposed $f_{is}$ contrast and neural soma density in specific regions of the brain cortex. The particular Brodmann parcellation available on FreeSurfer and used in this study does not contain all the areas identified by Broadmann in his seminal atlas (60). However, the main BA, characterized by distinctive differences in neural soma densities, such as 1-3 (somatosensory areas), 4 (primary motor area), 6 (pre-motor area), 17 (primary visual area), 18 (secondary visual area), 44 (Broca's area, pars opercularis) and 45 (Broca's area, pars triangularis) are represented with high fidelity, following a rigorous probabilistic parcellation procedure performed by Amunts and Zilles (61). For some of these areas, we provide also examples of histological images of cytoarchitecture from literature (62-64), showing differences in neural soma arrangement and density.

*4. Results*

*Regime of validity of the non-exchanging compartment model for different diffusion times and b values*. Results from the first numerical simulation experiment (Simulation 1) are reported in **Figure 3**. Specifically, **Figure 3a** shows three different simulated ADC dependences on diffusion time $t_d$: for the simulations where exchange between soma and neurites was considered ('exchange'); for the case of non-exchange ('no exchange') and the prediction of a simple compartment model (diffusion in randomly oriented sticks + GPD approximation of restricted diffusion in spheres) with the relative diffusivities, $f_{in}$, $f_{is}$ and $r_s$ known by construction ('compartments'). The difference between exchange and non-exchange conditions $\Delta$ADC as a function of $t_d$ are reported in **Figure 3b**, together with the threshold at 10% chosen as a reasonable level of approximation for modelling purposes. Considering different overall sizes of brain cell domains, ranging from 100 $\mu$m (approximating microglia) to 1000 $\mu$m (approximating big neurons), **Figure 3b** suggests that $t_d \leq 20$ ms is the diffusion time regime where the exchange between soma and neurites can be neglected and a simple compartment model can be used to model the overall intra-cellular signal as a sum of two non-exchanging compartments, namely intra-neurite and intra-soma. This regime of validity for different b values is further demonstrated by the results in **Figure 4** (Simulation 2), where the direction-averaged signal as a function of $b^{-1/2}$ is shown for two different $t_d$: $t_d$ = 10 ms < 20 ms (**Figure 4a**) and $t_d$ = 80 > 20 ms (**Figure 4b**). While signals for exchange, no exchange and simple compartments perfectly overlap in almost all the simulated scenarios at $t_d$ = 10 ms (**Figure 4a**), they are clearly different at $t_d$ = 80 ms (**Figure 4b**).

**Figure 3** suggests that $t_d \leq 20$ ms is a suitable threshold that on one hand offers sufficiently long diffusion time to probe soma structures of diameter up to ~24 $\mu$m, and on the other introduces only a small error (<<10%) when using our multi-compartment analytical model in Eq.(10). This is true for cellular structures similar to large neurons and medium-size neurons or glia (like astrocytes and oligodendrocytes) (first two rows in **Figure 3**). However, for the typical size of microglia-like cellular structures, such as the panel of $r_s$=4 µm and $f_{is}$=0.31 in **Figure 3**, the error expected from using the multi-compartment analytical model is too high (~20%) at $t_d$=20 ms. This suggests that the suitable $t_d$ must be chosen according to the desired sensitivity to specific cell domains. For example, to support the study of microglia-like structures, $t_d$<10 ms must be used. However, this choice would reduce the sensitivity to soma of bigger cell-types, such as big neurons (e.g. $r_s$=8 µm and $r_s$=10 µm columns) because the characteristic length scale probed at $t_d$<10 ms is less than 10 µm. **Figure 4** also shows the same effect, but through the b dependence of the direction-averaged normalized signal. At $t_d$=10 ms (**Figure 4a**) the analytical model well describes the signal attenuation as a function of b (up to very high b=60,000 s/mm$^2$, or equivalently 60 ms/µm$^2$), except for the case of very small cell domain and soma (panel $r_s$=4 µm and $f_{is}$=0.31 in **Figure 4a**). However, when longer diffusion time is used ($t_d$=80 ms, **Figure 4b**), the analytical model fails to describe the signal attenuation as a function of b also for larger cell domains and soma (second and third rows in **Figure 4b**). These results are confirmed by the direct comparison between ground-truth values and model-parameter estimates from fitting Eq.(10) without extra-cellular compartment to the simulated signals with exchange in **Figure 4**, and reported in **Figure 5**. The estimates for the case $t_d$ = 10 ms are consistently close (within the error) to the ground truth for almost all the scenarios, except for the small cell domain (third row in **Figure 4a**). In contrast, the estimates for the case $t_d$ = 80 ms are similar to or worse (within the error) than those at $t_d$ = 10 ms, especially for the small cell domains (second and third rows in **Figure 4b**) scenarios. From **Figure 5**, we also note that in some cases, such as $f_{is}$ = 0.30 in cell domain 1000 µm and $f_{is}$ = 0.32 in cell domain 400 µm, the estimates at $t_d$ = 80 ms are closer to the ground truth than those at $t_d$ 10 ms. However, in these cases the standard deviation on the estimated parameters (error bars in **Figure 5**) is higher, suggesting that higher fit instability and uncertainty in the parameter estimation may be the cause.

Finally, we note that these results do not change even if a much higher bulk diffusivity is used in our simulations, e.g. 3 µm$^2$/ms (see supplementary **Figure S1**).

*Sensitivity to soma size and density.* Results from Simulation 3 are reported in **Figure 6** where a dictionary of pre-computed direction-averaged simulated signals as a function of b$^{-1/2}$ is compared to real DW-MRI signal averaged across two ROIs representative of WM (SNR = 50, at b = 40 ms/µm$^2$) and GM (SNR = 20, at b = 40 ms/µm$^2$), collected in ex-vivo mouse brain. First, observe that in **Figure 6b**, only the cases marked with # closely mirror the experimental data at b>3 ms/µm$^2$ for WM, while only the cases marked with * for GM. The first set of cases corresponds to the cellular configurations $f_{is}$ ~ 1-5% and $r_s$ = 2 µm in **Figure 6a**, while the second set corresponds to $f_{is}$ ~ 60-65% and $r_s$ = 6-10 µm. These configurations match our understanding of the neuroanatomy: in the WM ROI (corpus callosum) only a small volume fraction is occupied by oligodendrocytes, whose elongated soma has shorter axis of only a few µm; while in GM ROI (cortex) there is an abundance of large soma with typical size ($r_s$ = 6-10 µm) compatible with cortical pyramidal neurons. Second, the other cases demonstrate the specificity of our computational model: different cellular configurations can produce a range of distinct signal variations. For example, the panels (first row, second column) and (third row, third column) in **Figure 6b** exhibit signal variations distinct from those of the two ROIs investigated. The corresponding panels in **Figure 6a** show that they correspond to distinct cellular configurations: one containing large cellular domains with low volume fraction of soma of intermediate size, while the other containing small cellular domains with very high volume fraction of soma of large size. It is worth noting that these results also suggest that SANDI performs fairly well for characterizing WM, which is consistent with the standard "stick" model being a valid approximation for WM (15). Furthermore, we see that for b-values as low as 3,000 s/mm² (or 3 ms/µm²), the contribution from extra-cellular water is negligible, as it is not present in our simulation but present in the experimental data.

*Accuracy and precision of model estimates*. The study of accuracy and precision of model-parameter estimates ($f_{is}$, $D_{in}$ and $r_s$) using the RF regression is reported in **Figure 7**, showing that the proposed model can closely approximate (within 10% bias, or 90% accuracy) the connected cellular structure in the ideal case (SNR=∞), and maintains good accuracy and precision in more realistic case of SNR = 50 and acceptable accuracy and precision in the worse-case scenario SNR=10.

The ablation study is reported in **Figure 8**, showing that a minimum of five b values (or b shells), with two of them higher than 3,000 s/mm² (or equivalently 3 ms/µm²) are required to produce parameter estimates of reasonable accuracy and precision. These results suggest that the in vivo human dataset used in this work (MGH Adult Diffusion Dataset) is adequate (third column in **Figure 8**), but less and/or lower b values would be insufficient, which would lead to MSE values 2 to 30 times larger (first and second column in **Figure 7**).

*Model parameters maps in human brain*. Parametric maps of all the five model parameters $f_{in}$, $f_{ec}$, $D_{in}$, $D_{ec}$, $r_s$ and $f_{is}$ for a representative human subject are reported in **Figure 9**.

*Comparison with histology*. Qualitative comparison of $f_{in}$ and $f_{is}$ maps with histological images (from different subjects) of myelo- and cyto-architecture (myelin- and Nissl-staining) from available human brain atlas (https://msu.edu/~brains/brains/human/) are reported in **Figure 10**.

*Comparison of the soma signal fraction map and brain cytoarchitecture in 25 healthy human subjects*. Examples of $f_{is}$ maps projected onto the cortical surface (representing the whole cortical thickness) of 4 representative subjects together with the average map over the 25 subjects analysed in this study are reported in **Figure 11**. Boundaries of Brodmann areas BA 1-6, 17, 18 and 44, 45 are also reported to show the remarkable match with boundaries where $f_{is}$ values change, in both individual subjects and average maps. Brodmann areas are known to identify regions of the brain cortex where cytoarchitecture differs in soma density and arrangement. **Figure 12** compares average MR estimates of soma density ($f_{is}$ map) among the 25 subjects with histological images from a few representative areas from literature showing a remarkable match of gradients in $f_{is}$ values with the gradient in soma density, one of the criterion used to parcellate the cortex in different Brodmann areas.

*Comparison with the dot-compartment variant*. In **Figure 13** we show the results of fitting SANDI model (Eq.(10)) and the dot-compartment variant (Eq.(12)) to the ex-vivo mouse data in Figure 6, for both WM and GM ROIs. The values of the estimated fitting parameters are reported in the table, together with the AICc values for each model. We found that SANDI describes the data better than the dot-compartment variant, wich lower AICc (ΔAICc>2), for both WM and GM.

## 5. Discussion

In summary, this work proposes SANDI, a novel model to estimate apparent soma and neurite density non-invasively using DW-MRI. Our approach challenges the existing standard model (2-8) that considers water diffusion in WM and GM as restricted diffusion in neurites, modelled by "sticks" embedded in the hindered extra-cellular water (**Figure 1a**). Motivated by recent studies that suggest this "stick" assumption fails in GM (10-12), we hypothesise that one plausible explanation for such failure is the abundance of cell bodies (namely soma) in GM relative to WM (12). So far, the contribution from soma has not been directly modelled, but rather assumed to contribute to the overall extra-cellular compartment (for example see (2, 38)) (**Figure 1a**). Indeed, the underlying assumption has been that there is negligible restriction in soma because of the fast exchange rate with the extra-cellular space. However, a recent estimate of such exchange reports a water pre-exchange lifetime in neurons and astrocytes > 500 ms (29). This suggests that for relatively short diffusion times $t_d$ << 500 ms, restriction in soma may be not negligible. Here, we use advanced numerical simulations in realistically connected cellular structures to show that soma has indeed a specific signature on the DW-MRI signal at high b-values. The results from ex-vivo experiments in a mouse brain show that the signature predicted in simulation is both present and observable in measured signals (**Figure 6**). The results from in-vivo experiments in a cohort of 25 healthy human subjects show that the proposed technique can provide maps of apparent soma density and size that meet expectations from histological imaging (**Figure 10**) and current anatomical understanding (**Figure 12**). These findings are also in agreement with other recent works that have challenged the validity of the standard model and its derived variants (like spherical mean technique (5)), and showed that factors not considered by the underlying microstructural models, such as intercomponent and intracompartmental kurtosis, may cause misestimation of the model parameters (13, 14).

*SANDI as a first model for soma imaging*. We propose a new microstructure model based on three non-exchanging compartments that explicitly includes the soma contribution to the intra-cellular signal as a pool of water diffusing in restricted geometries of non-zero size, i.e. not a dot-compartment (9, 11, 65), but rather a restricted water pool, whose MR signal has a specific b and $t_d$ dependence (i.e. Eq. (8)) (**Figure 1b**). Our numerical simulations (**Figure 3-5**) identify the time regime of validity for such a simple compartment model to be at relatively short diffusion times ($t_d \leq 20$ ms). In this time regime, the exchange between intra- and extra-cellular compartments is also negligible, for both neuronal and glial cell types, as suggested by a recent study from Yang et al. (29). Furthermore, we show with numerical simulations and experiments in mouse brain that soma size and density have indeed a specific signature on the direction-averaged DW-MRI signal at high b values (i.e. $b_{max} > 3,000$ s/mm$^2$, corresponding to 3 ms/μm$^2$) and that the high b value regime can be used to increase sensitivity to geometrical restrictions of typical neural soma size ranging from a few microns (e.g. for microglia and glia) to a few tens of microns (e.g. for big neurons) (66, 67) (**Figure 4-6** and **8**). In the simulations we performed in this work, we did not include long axons because they have been shown to have small impact on the measured intra-cellular diffusion (see supplementary information of (39)). Under these experimental conditions, the normalized direction-averaged (or powder averaged) DW-MRI signal can be expressed as the sum of three non-exchanging compartments: intra-cellular, comprised of intra-neurite and intra-soma compartments, and extra-cellular compartment (**Figure 1b**). We show that such a model, under such experimental conditions, approximates very well the expected intra-cellular signal (**Figure 3-6**) and provides reasonably accurate and precise estimates of neurite MR signal fraction and soma MR signal fraction and apparent size (**Figure 5, 7** and **8**). Note that the proposed model is very different from previously proposed models that include a sphere compartment to account for other extra-neurite compartments, such as in Stanisz et al. (32). In fact, here we propose a model to disentangle the signal from cell-bodies of any cell type (modelled as a sphere compartment), from the signal from elongated cellular projections, such as neuroglial processes and neuronal dendrites and axons (modelled as sticks). In contrast, Stanisz et al. (32) proposed a sphere compartment to characterize the signal coming from the whole cell domain (hence cell body + cellular projections) of glial cells in bovine optic nerve.

*Comparison with dot-compartment variant*. The results reported in **Figure 13** show that SANDI describes the measured MRI signal better (lower AICc) than the model comprising a dot-compartment (ΔAICc>2), in both GM and WM. More importantly, we found that the signal fractions estimated using the dot-compartment model do not correspond to the neuroanatomy expected from histology and the apparent diffusivities are not in agreement with literature, especially for GM data. For WM, the dot-compartment model and SANDI provide similar estimates, suggesting that for WM, which has low soma density, both models are good approximations (15), with SANDI still providing better fit (lower AICc).

*Non-invasive cyto- and myelo-architecture maps by DW-MRI*. Using the newly introduced microstructure model, we retrospectively analysed data from 25 healthy subjects from the MGH Adult Diffusion Dataset, freely available from the HCP data repository. This dataset was coincidently acquired with experimental conditions appropriate for our model assumptions: $t_d \leq 20$ ms and $b_{max}$ = 10 ms/μm$^2$. Parametric maps of the estimated model parameters in **Figure 9** show very reasonable and encouraging contrasts. Maps of $f_{in}$ have contrast highlighting the major WM tracts in the brain while $f_{is}$ values are higher in GM (**Figure 9**). Specifically, $f_{in}$ values are higher in voxels mostly comprised of WM and they seem minimally affected by fibre orientation dispersion and crossing, as shown by the uniform contrast in all the WM regions, even in those characterized by high fibre orientation dispersion and crossing, e.g. regions where the radiation of the corpus callosum and the corona radiata cross. We also notice that $f_{in}$ values in cortical GM are between ~0.1-0.2, in good agreement with recent works focusing on estimating neurite density in GM using more advanced DW-MRI acquisition schemes such as spherical tensor encoding (19, 20). Values of $f_{is}$ are consistently higher in all GM regions, from cortical to deep and cerebellar GM. The slightly lower values of $f_{is}$ in cerebellum may be due to higher partial volume in this region between WM and GM due to the large size of MRI voxels (1.5x1.5x1.5 mm$^3$). We note that more advanced DW-MRI acquisition schemes such as B-tensor encoding (68) seem to offer encouraging preliminary results on $f_{is}$ estimation using lower b values (69), in good agreement with our estimates. Values of $r_s$ across the brain range from 2 to 12 μm, with a mean±std values of 10±3 μm. These values are in good agreement with the expected mean±std radius of neural soma in human brain 11±7 μm, as evaluated by a supplementary analysis we performed using about 3000 reconstructions of human brain cellular morphologies, available from the Neuromorpho database (neuromorpho.org). Finally, $D_{in}$ and $D_{ec}$ values are in good agreement with published values for intra-neurite diffusivity in WM being about 2.3 μm²/ms (70) and extra-cellular diffusivity $D_{ec} < D_{in}$ (71). However, given the non-optimal experimental design of the human dataset, the maps of $r_s$ should be taken with care. In fact, $r_s$ estimation in this case is neither very accurate nor particularly precise because of the limited number of b values and only one t_d. This is shown in supplementary **Figure S2**, where the random forest regressor predicts different and not always sensible values for $r_s$ when trained with an incorrect range, e.g. $r_s$=[1, 20] μm, instead of the correct one ($r_s$=[1, 12] μm). The model parameters $f_{is}$ and $f_{in}$ are MR signal fractions of the two compartments that in our model are linked to soma and neurites. As such, we expect them to correlate with soma and neurite densities or volume fractions within the MRI voxel. On the other hand, Nissl and myelin are two of the most used staining to characterize the brain cyto- and myelo-architecture. Although Nissl staining stains mostly the nucleus of neural cells rather than the whole cell body, we can reasonably expect that the contrast in Nissl staining correlates with density and arrangement of soma and thus the cyto-architecture of the brain. Similarly, myelin staining stains only the myelinated neurites, such as axons or myelinated dendrites, thus we can reasonably expect that the contrast in myelin staining highlights mostly WM tracts and the myelo-architecture of the brain. The qualitative comparison of $f_{is}$ and $f_{in}$ maps with Nissl- and myelin-stained histological images in **Figure 10** show a remarkable similarity between the MRI maps and the contrast in the histological images, suggesting that $f_{is}$ and $f_{in}$ maps could be used to characterize in a non-invasive way the cyto- and myelo-architecture of neural tissue. However, we note that the concordance of SANDI maps and histology is not perfect. This could be due to several reasons: implicit contribution to f_{is} from relaxation weighting; unavoidable differences between histology and imaging (e.g. thinner slice and higher in plane resolution in histology); histological and MRI images from different subjects. A particularly important difference is that, unlike its histological counterpart, f_{is} is a fraction, thus reflecting the signal contribution of soma relative to the entire intra-cellular space. For instance, we note, unlike the Nissl staining image, f_{is} values in the thalamus are lower than other neighbouring GM regions (e.g putamen and caudate). This is consistent with the fact that the thalamus, different from its neighbouring GM structures, consists of a large amount of WM and myelin (72-74) (e.g. the stratum zonale that covers the dorsal surface and the external and internal medullary laminae), leading to lower f_{is} values (and higher f_{in} correspondingly). Moreover, histological images and fitted SANDI parameter maps are from different subjects. On one hand, this represents a limitation of the present work that future validation work (75) will aim to address. On the other, it makes the observed concordance between SANDI parameter maps and histology remarkable.

*Non-invasive cyto-architectonics of the human brain by DW-MRI*. Although a proper validation of the novel contrast in soma density and size introduced by the new model presented here is still in progress (75), here we present individual and average results over 25 healthy subjects, showing remarkable similarity of $f_{is}$ values distribution on the cortical surface with Brodmann areas parcellation of the brain cyto-architecture (**Figure 11** and **12**). As shown in **Figure 12**, Brodmann areas are characterized by different soma density and arrangement. Changes in $f_{is}$ values on the cortical surface (representing the whole cortical thickness) follow very well the boundaries of Brodmann areas (within acceptable slight mis-alignment probably due to small errors in the co-registration procedure, performed using the automated toolbox within FreeSurfer), demonstrating that the contrast provided by this new MRI parameter could be used as non-invasive imaging marker of cyto-architectonics. Furthermore, the correspondence to Brodmann areas is also very good at the level of individual subject, as shown by **Figure 11**.

*Potential impact*. A deeper understanding of cortical organization, including its complex fiber architecture and structural connectivity is still an open challenge in neuroscience. There has been considerable interest in the mapping of GM microstructure. Some examples include cortical laminar structure characterization using high-resolution DTI (76), assessment of GM maturation in rodents with DKI (77), and neurite density and dispersion quantification in human brain with NODDI (78-80), all using PGSE experiments at intermediate diffusion times. Other works have used PGSE experiments at shorter diffusion times ($t_d \leq 30$ ms) (81, 82), or oscillating gradients at ultra-short diffusion times (83), to characterize the GM microstructure. SANDI can help interpret and model in terms of soma size and density their results concerning, for example, DIAMOND (84) and restriction spectrum imaging (RSI) (81) metrics, and ADC frequency dependence in GM (83). The additional contrasts provided by SANDI maps of soma MR signal fraction and apparent MR measured soma size could be useful to develop novel quantitative cytoarchitectonic and probabilistic mapping of cortical areas in a whole-brain fashion. For example, these could be used to define new parcellations of the brain based on cytoarchitectural features (e.g. **Figure 10** could be a proof-of-concept of it); and to improve the quality of currently available atlases of brain GM sub-regions that are notoriously difficult to delineate, such us the numerous nuclei comprising the brainstem (85) (e.g. some encouraging preliminary results have been recently shown in (86)). SANDI could also help provide metrics more specific to changes in the brain cyto- and myelo-architecture through development and due to the onset of disease. In longitudinal studies of developing brain (87-95), Ouyang et al. (95) has recently used DKI (96) to quantify the dynamic cortical microstructural signature of critical developmental stages. SANDI may help understand the exact neuroanatomical underpinning of observed cortical MK in terms of changes in neuronal soma density. In Multiple Sclerosis (MS), SANDI could provide more specific information about microglial and astrocytic activation during inflammation and astrocytic scarring, both processes involving increased accumulation of glial soma within the MS lesions (97) (e.g. some encouraging preliminary results have been recently shown in (98)). In epilepsy, SANDI maps of apparent MR measured soma size and soma MR signal fraction could improve sensitivity and specificity to the remodelling of brain cytoarchitectonic occurring within the epileptic lesions (99).

*Data acquisition requirement*. In this work a comprehensive scheme was used a-priori so as to ensure high accuracy and precision of our results (**Figures 7** and **8**). Further refinement and optimization will be required in the future to establish the limits of the methodology. In general, given the specific hardware characteristics of an MRI scanner (either preclinical or clinical) and acquisition time constraints, it is always possible to optimize the experimental protocol in order to achieve reasonable levels of precision in SANDI model parameters estimation. A possible general approach to perform such optimization has been previously proposed by Alexander (100).

It is important to underline that conventional clinical DW-MRI data are in general not suitable for SANDI, for a number of reasons. First, b values are typically no higher than 3,000 s/mm$^2$, or equivalently 3 ms/μm$^2$. SANDI requires several much higher b values. Second, diffusion times are typically much longer than 20 ms, the upper bound we have identified for SANDI modelling to be valid. The ablation study demonstrates that the MGH Adult Diffusion Dataset used in this work is an example of a minimal dataset that meets these requirements: short enough diffusion time, high enough b values, six b-shells to estimate five model parameters.

Similar considerations hold for ex-vivo experiments but as these experiments are commonly performed at room temperature, we must take into account the resulting lower diffusivity. Typically, apparent diffusion coefficient in ex-vivo brain at room temperature of 21°C can be ~4 times lower than in-vivo, suggesting that longer time scales (e.g. $t_d$≤80 ms) and higher b values (e.g. >12,000 s/mm$^2$, or equivalently 12 ms/μm$^2$) have to be used. The ex-vivo mouse brain dataset used in this study is an example of a suitable ex-vivo experimental protocol.

*Future directions and perspectives*. The SANDI model presented here for apparent soma size and density estimation is used to analyse DW-MRI data acquired with simple classical MRI acquisition schemes like PGSE or Pulsed Gradient Stimulated Echo sequences. However, other more advanced acquisition schemes such as double diffusion encoding (DDE) or B-tensor encoding, may provide improved accuracy and precision of SANDI model parameters estimation. In fact, it has been shown that these acquisitions provide additional information that can help disentangling the relative contribution of the different compartments modelled by SANDI. For instance, B-tensor encoding has been successfully used to improve the estimation of neurite density based on the standard model (3) (19, 20), while recent works using DDE showed that this acquisition scheme can help disentangling different sources of DW-MRI signal that can be linked to different features of the underpinning tissue microstructure (101-104). Future works will focus on harnessing the orthogonal information offered by these advanced acquisition schemes in order to maximize the sensitivity and specificity of the measured DW-MRI signal to the soma contribution (e.g. some encouraging preliminary results have been recently shown in (105)). Together with improving the acquisition, another target of future work is the rigorous validation of the new parametric maps provided by SANDI. As already mentioned, SANDI provides apparent soma size and density maps in terms of MR measured spherical compartment size and relative signal fraction, respectively. By model design, these values are expected to correlate with the higher moments of the actual soma size distribution and the soma density in the brain tissue. However, histological validation is complex, expensive, and time consuming. Histological measurements of cellular content have their own inaccuracies and inconsistencies and appropriate metrics are difficult to fine tune. Preliminary investigation in ex-vivo mouse brain at ultra-high field and comparison with histological staining for cell bodies has already shown encouraging strong positive correlations between the soma size and density estimated from SANDI and those directly measured from histology (75). Future work will extend such investigation to different mouse brain regions, such as cerebellum and olfactory bulb, and will provide more quantitative proof of the actual link between SANDI model parameter and the actual brain tissue microstructure. Nevertheless, the qualitative results (**Figure 10** and **12**) showing trends consistent with known histological variation at macroscopic scale support the first demonstration of the value of SANDI.

## 6. *Conclusion*

The current conjecture in brain microstructure imaging (2-9) envisions the brain tissue component in an MRI voxel as subdivided into two non-exchanging compartments: intra-neurite and extra-neurite space. The total MRI signal is then given by the weighted sum of the signals from water molecules diffusing in each compartment. Although very successful in describing the DW-MRI signal in WM and GM at relatively low b values (b≤3,000 s/mm$^2$, or 3 ms/$\mu$m$^2$) in both healthy and diseased conditions (2-8, 18-26), this microstructure model fails in describing DW-MRI signal at high b values (b>>3,000 s/mm$^2$, or 3 ms/$\mu$m$^2$) (10-12, 15). Here we introduce a new picture: the tissue component of an MRI voxel is subdivided into intra-cellular and extra-cellular non-exchanging compartments. The total signal is the weighted sum of the signal from water molecules diffusing in each compartment. Furthermore, the intra-cellular compartment is itself divided into two non-exchanging sub-compartments: intra-neurite and intra-soma. The intra-cellular MRI signal is then given by the weighted sum of the MRI signal from water molecules diffusing within the two sub-compartments. This new microstructure model that we call SANDI (Soma And Neurite Density Imaging) directly accounts for one of the major differences between WM and GM: soma abundance in GM compared to WM, enabling the non-invasive estimation of apparent soma density and size trough MRI.

Using advanced numerical simulations, we identified the regime of validity of the assumption of non-exchanging intra-cellular sub-compartments (neurites and soma) and propose SANDI as a new method for non-invasive soma imaging. We demonstrate it in ex-vivo DW-MRI mouse data and in-vivo cutting-edge human acquisitions. We showed that the new microstructure model for soma imaging defines new contrasts, dissimilar to the simple tensor analyses, representing new complementary information on the brain cyto and myeloarchitecture. Although still under validation (75), the maps here reported already show some interesting contrast that might provide new insight into tissue architecture and provide markers of pathology, as well as a new set of biomarkers of potential great value for biomedical applications and pure neuroscience. With the availability of powerful human scanners like the Connectom (58), this technique has the potential for translation into the clinic, opening a promising avenue for more in-depth assessment of cellular microstructure *in-vivo* in human brain.

**Acknowledgments**

This work was supported by EPSRC (EP/G007748, EP/I027084/01, EP/L022680/1, EP/M020533/1, EP/N018702/1, EP/M507970/1 and European Research Council (ERC) under the European Union's Horizon 2020 research and innovation programme (Starting Grant, agreement No. 679058).

**Figures**

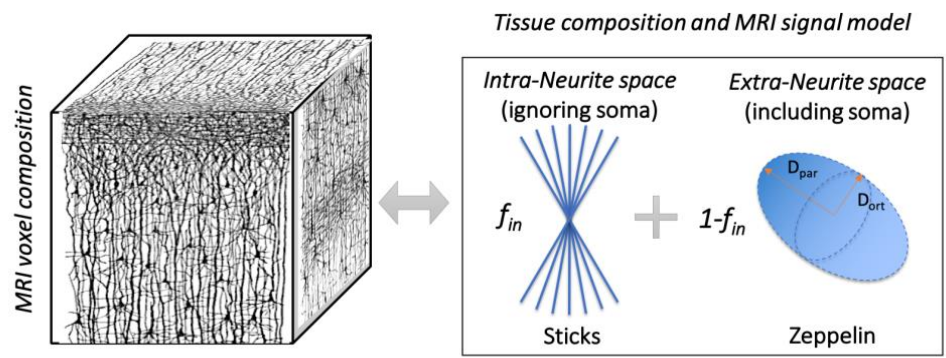

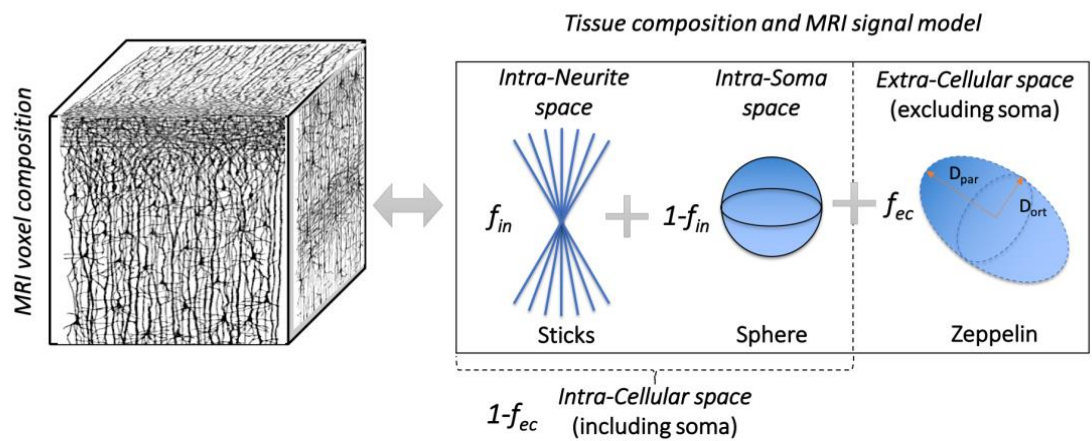

*Figure 1 Schematics of current standard model of brain microstructure (a) and the novel model proposed in this work (b). Current conjecture envisions the tissue component in an MRI voxel as subdivided into two non-exchanging compartments: intra-neurite and extra-neurite space. The total MRI signal is then given by the weighted sum of the signals from water molecules diffusing in each compartment, with relative signal fractions $f_{in}$ and $1-f_{in}$, respectively (a). We propose a new picture: the tissue component of an MRI voxel is subdivided into intra-cellular and extra-cellular non-exchanging compartments. The total signal is the weighted sum of the signal from water molecules diffusing in each compartment, with relative signal fractions $1-f_{ec}$ and $f_{ec}$, respectively. Furthermore, the intra-cellular compartment is itself divided into two non-exchanging sub-compartments: intra-neurite and intra-soma. The intra-cellular MRI signal is then given by the weighted sum of the MRI signal from water molecules diffusing within the two sub-compartments, with relative signal fractions $f_{in}$ and $1-f_{in}$, respectively (b).*

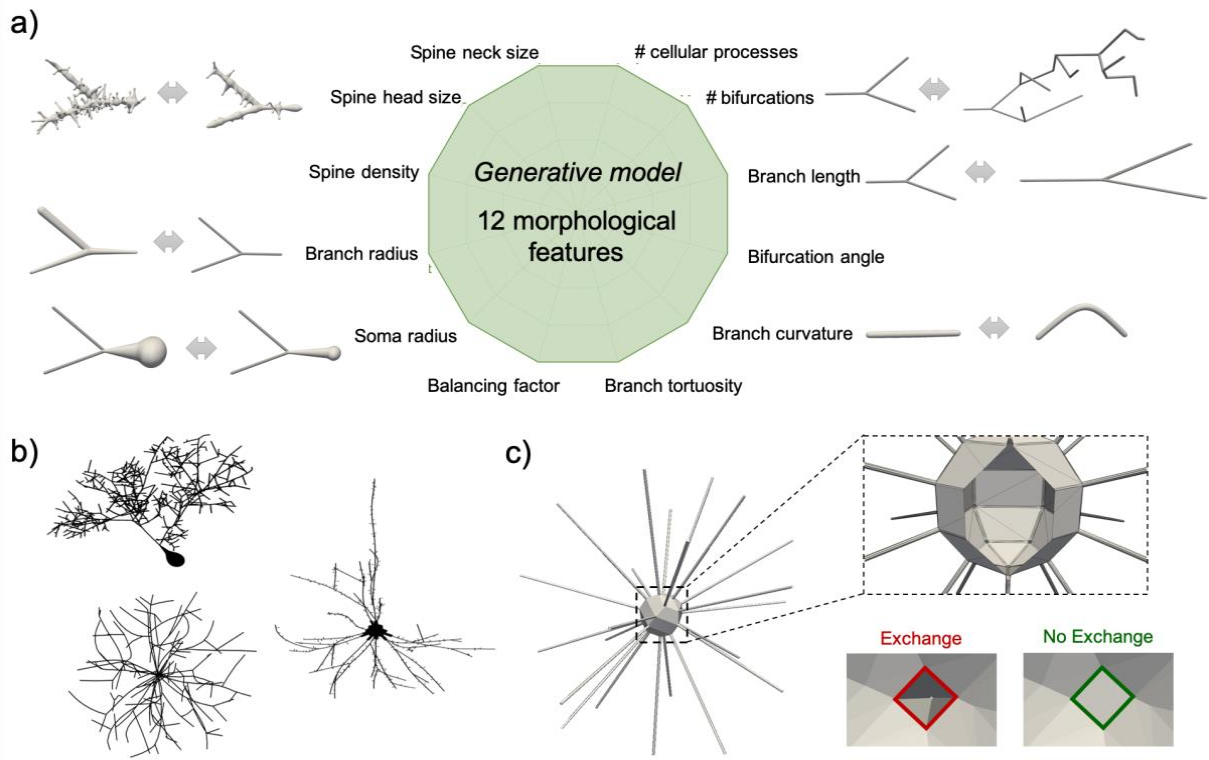

***Figure 2*** *Summary and a few examples (a) of the 12 morphological features used in the generative model of brain cells generation introduced in (1) to simulate realistic cellular structures like Purkinje cells, motor neurons and pyramidal spiny neurons (b). Here, the generative model is used to investigate simplified cellular structures (c) comprised of straight long cylindrical fibres connected to a spherical soma structure, with and without the possibility for diffusing spins to exchange between neurites and soma.*

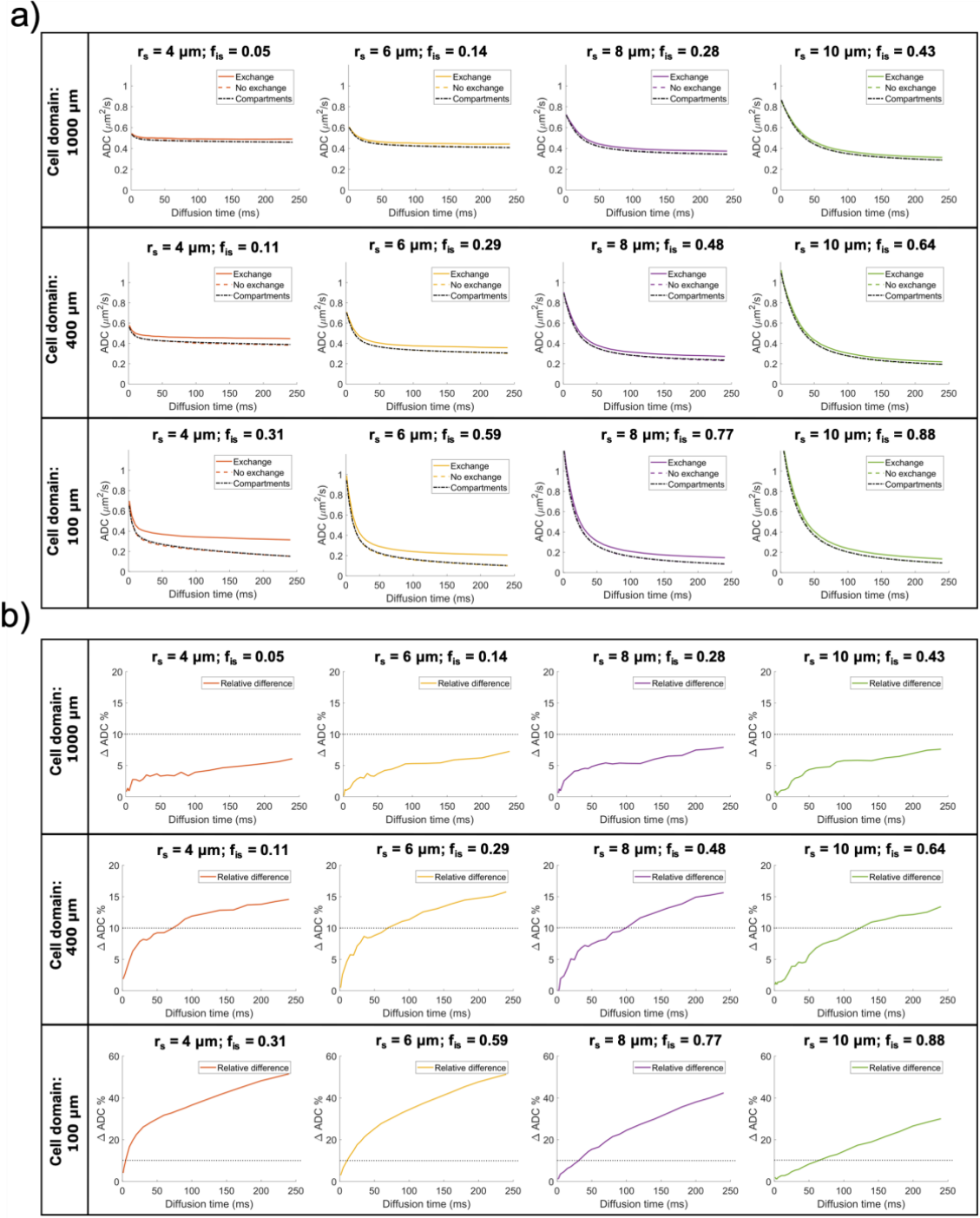

***Figure 3*** *Regime of validity of the compartment model. a) Comparison of the diffusion time dependence of the apparent diffusion coefficient (ADC) in cellular structures of different overall size and soma size/density, for three conditions: 1) fully connected cellular structures, simulating exchange between soma and neurites (exchange); 2) cellular structures where the connections between soma and neurites have been closed, simulating no exchange between soma and neurites (no exchange); 3) ADC computed from the compartment model (10) in the GPD approximation, without extra-cellular compartment (compartments). b) relative percentage difference between the ADC in the exchange and no exchange cases in a). The dashed lines show the 10% threshold used to define the diffusion time regime where the compartment model (10) is a reasonable approximation of cellular structures.*

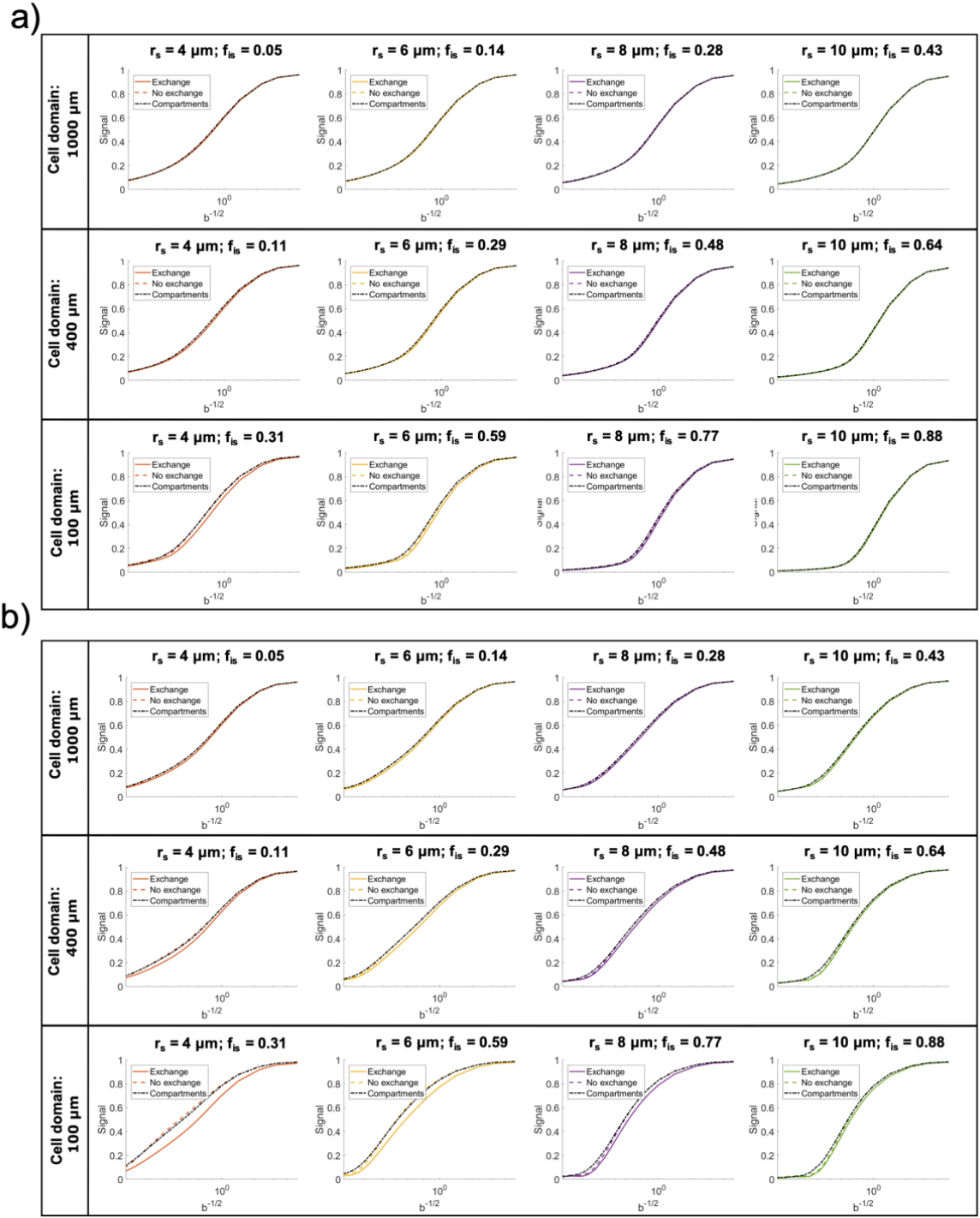

***Figure 4*** *Direction-averaged normalized signal as a function of $b^{-1/2}$ in $(ms/\mu m^2)^{-1/2}$ for two diffusion times: 10 ms, where, according to the results in **Figure 3**, we expect the compartment model to be a good approximation of the intra-cellular signal (a) and 80 ms where we expect the compartment model to fail (b). As in **Figure 3**, cellular structures of different overall size and soma size/density were considered, for three conditions: 1) exchange allowed between soma and neurites (exchange); 2) exchange not allowed between soma and neurites (no exchange); 3) computed from the compartment model (10) in the GPD approximation, without extra-cellular compartment (compartments).*

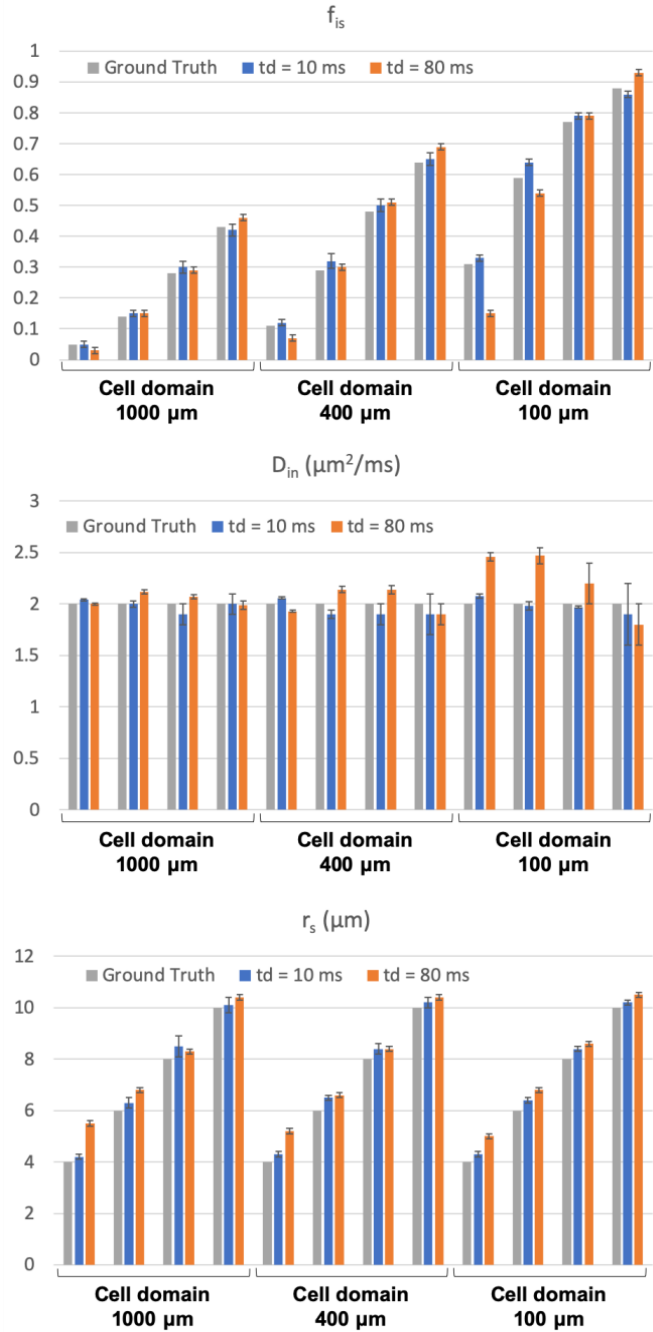

*Figure 5* Comparison between ground-truth model parameters and their estimates from fitting Eq.(10) without extra-cellular compartment to the simulated signals with exchange between neurites and soma in **Figure 4**. Uncertainty (error bars) in parameter estimates was quantified by a Monte Carlo approach (see Methods section 3.1 for details).

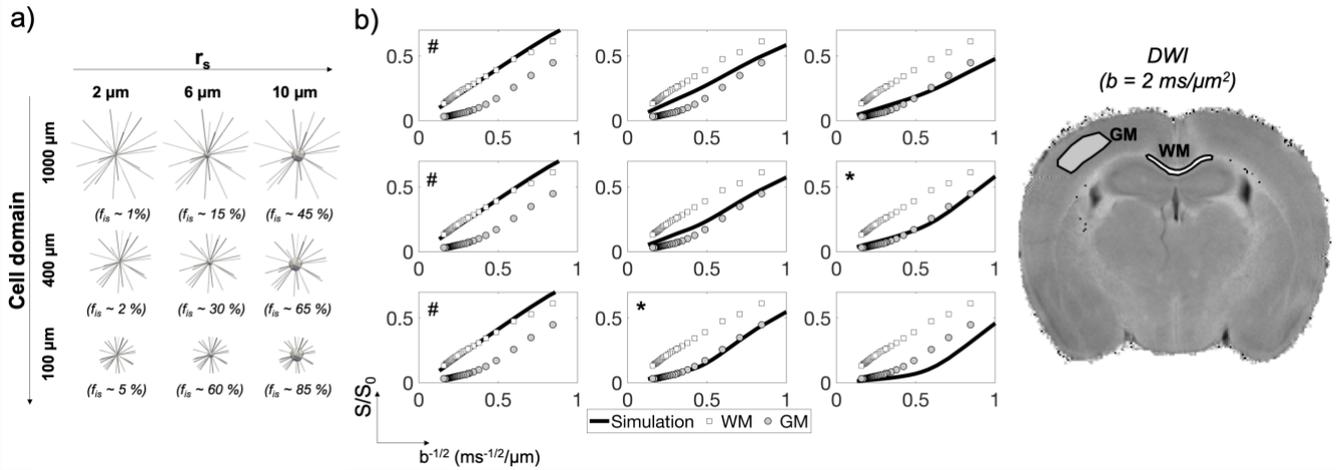

*Figure 6* Comparison of numerical simulations with ex-vivo mouse data. a) Computational model: many randomly oriented cylindrical segments were projected from a spherical compartment (soma) of radius $r_s$, at different volume fractions of soma $f_{is}$ for different overall cell size (cell diameters). b) Normalised direction-averaged DW-MRI signal as a function of $b^{-1/2}$ computed from spin-trajectories simulated in the structures in a) (line). Comparison with measured signal from white matter (WM) and gray matter (GM) ROIs shows very good match at $0.2 < b^{-1/2} < 0.5$ $(ms/\mu m^2)^{-1/2}$ for $(r_s, f_{is})$ conditions: (2, 0.01-0.05)WM marked as # and (6-10, 0.5-0.6)GM marked as *. Comparison with simulation also suggesting negligible extra-cellular contribution for $b^{-1/2} < 0.6$ $(ms/\mu m^2)^{-1/2}$, corresponding to $b \geq 3$ $ms/\mu m^2$.

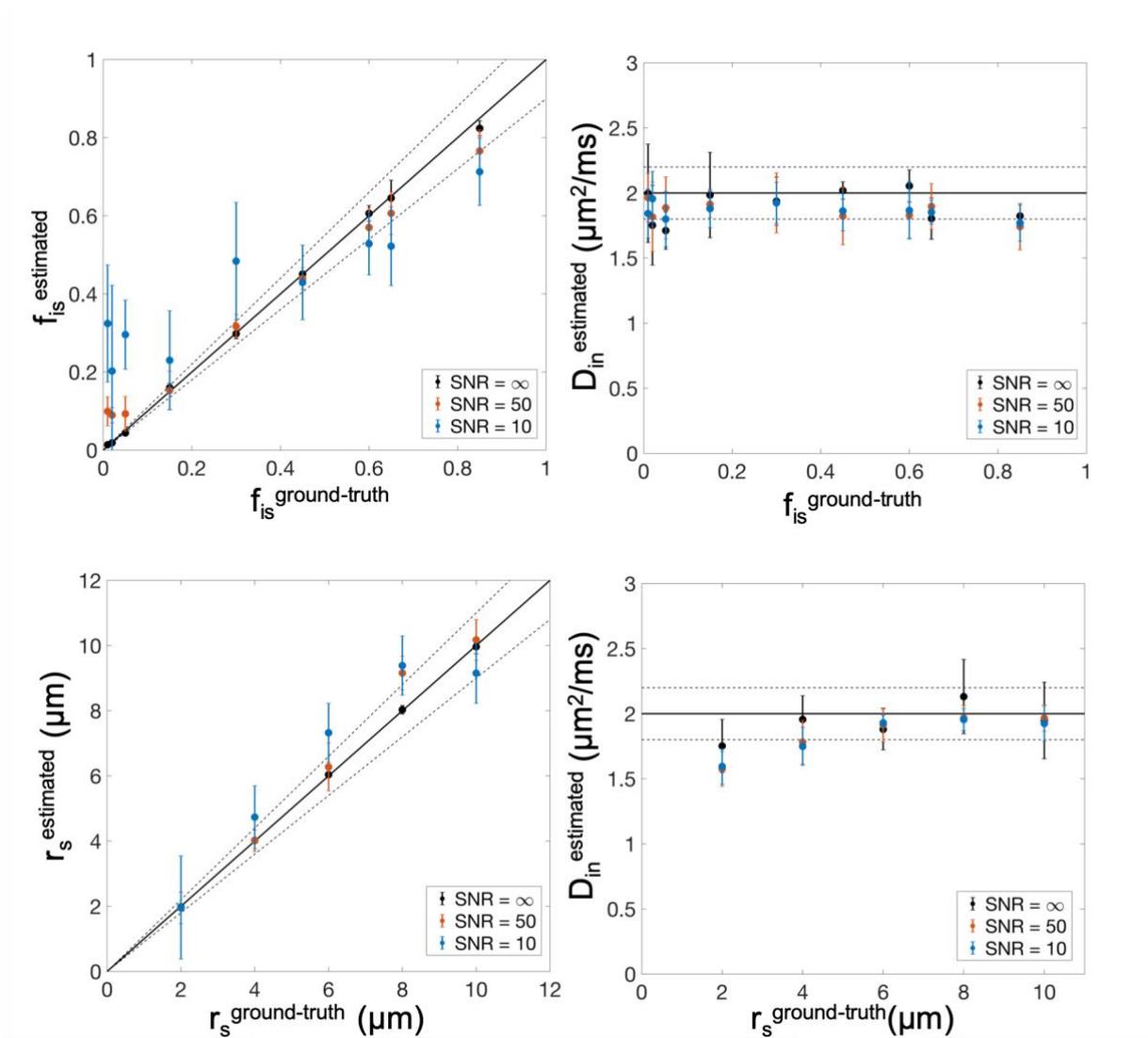

*Figure 7* Correlation accuracy plot. Soma compartment signal fraction $f_{is}$, soma apparent size $r_s$ and axial intra-neurite diffusivity $D_{in}$ estimated using relation (10) without the extra-cellular compartment and GPD approximation and labelled with the superscript "estimated" are plotted against the ground truth values labelled with the superscript "ground-truth". The perfect positive correlation line (solid line) and ±10% error (dashed lines) are plotted. In infinite SNR case, the correlation is very high ($R^2>0.98$) and bias within 10%. In the more realistic scenario of SNR = 50 the correlation is still very high ($R^2>0.85$) and accuracy and precision are close to the ideal case of infinite SNR. In the worse-case scenario of SNR = 10, the correlation is still high ($R^2>0.75$) and accuracy and precision acceptable. Error bars on data points indicate the statistical dispersion (standard deviation) in model parameter estimation as evaluated by Monte Carlo approach (2500 random drawn).

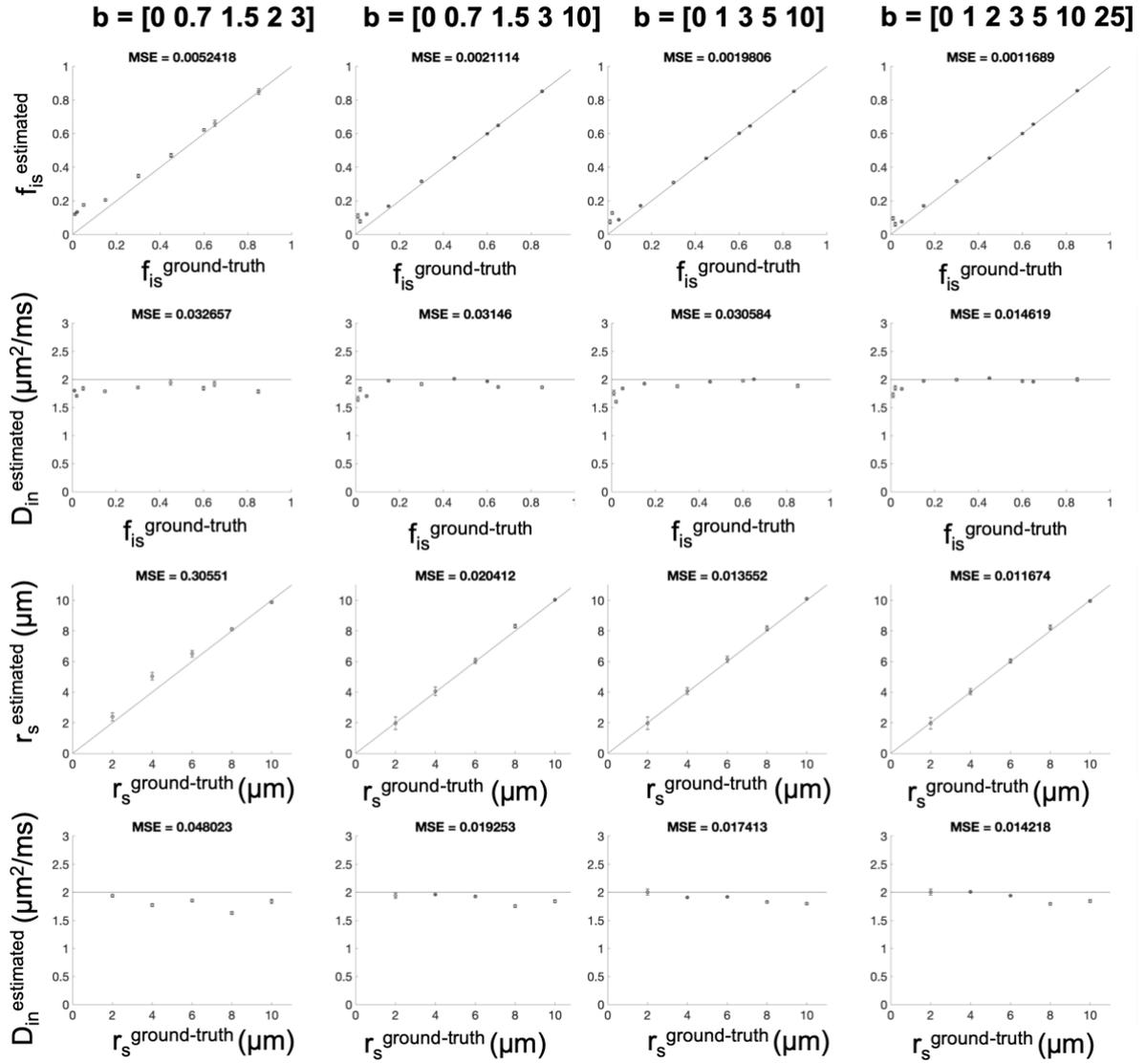

*Figure 8 Ablation study of accuracy and precision of model parameters estimation. Soma compartment signal fraction $f_{is}$, soma apparent size $r_s$ and axial intra-neurite diffusivity $D_{in}$ estimated using relation (10) without the extra-cellular compartment and GPD approximation and labelled with the superscript "estimated" are plotted against the ground truth values labelled with the superscript "ground-truth" for four different b value combinations (subsampled from Experiment 2 with $\delta/\Delta$ = 3/11 ms). The perfect positive correlation line is plotted as solid line. Error bars on data points indicate the statistical dispersion (standard deviation) in model parameter estimation as evaluated by Monte Carlo approach (2500 random drawn) in case of infinite SNR. The mean squared errors (MSE) with respect to the ground-truth values are reported for each b value combination as metrics of accuracy and precision (lower the value, higher the accuracy and precision).*

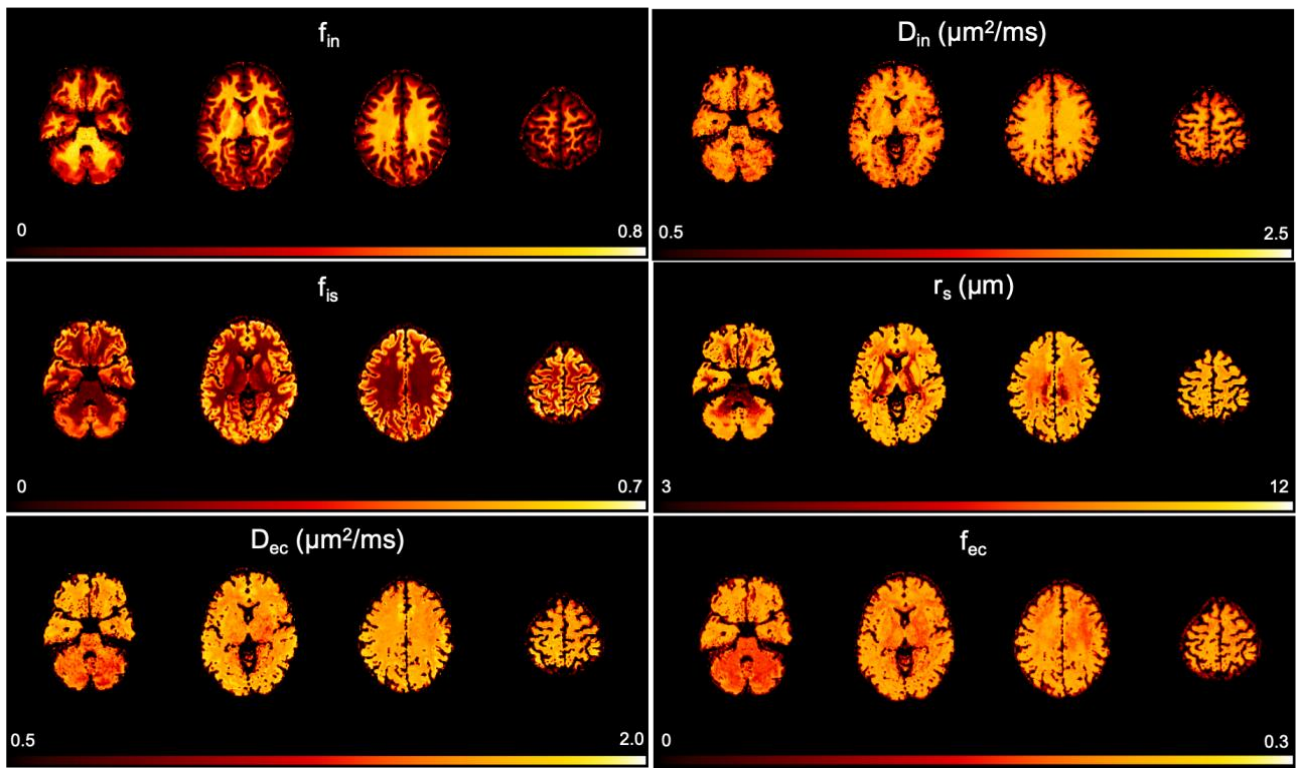

***Figure 9*** *An example of the parametric maps of the proposed compartment model for brain microstructure, obtained by fitting Eq. (10) to the normalized direction-averaged DW-MRI data from a representative subject.*

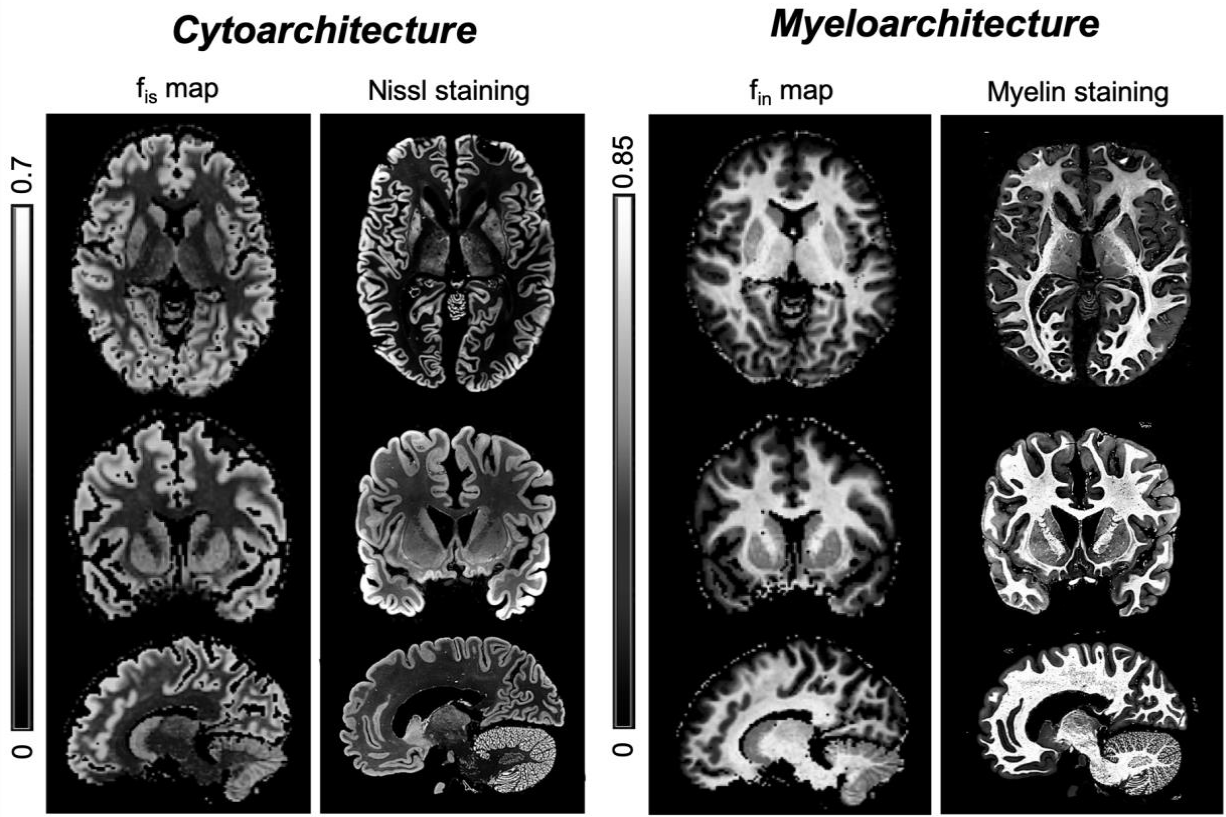

*Figure 10* *Novel contrasts in apparent neurite and soma density of human brain. Qualitative comparison of MR soma signal fraction $f_{is}$ and MR cell fibers signal fraction $f_{in}$ maps with histological images (from different subjects) stained for brain cytoarchitectonic (Nissl staining for cell nuclei, left side) and myeloarchitectonic (myelin staining, right side), respectively. Brain histological images from the human brain atlas at https://msu.edu/~brains/brains/human. The contrast in the MRI maps show remarkable similarity to the contrast from histological staining. Quantitative map of $f_{is}$ is expected to provide contrast related to soma density, while map of $f_{in}$ is expected to provide contrast related to neurite density. Since myelinated neurites (like axons) are the major constituent of white matter, $f_{in}$ is expected to provide contrast highly related to myelin in the white matter regions.*

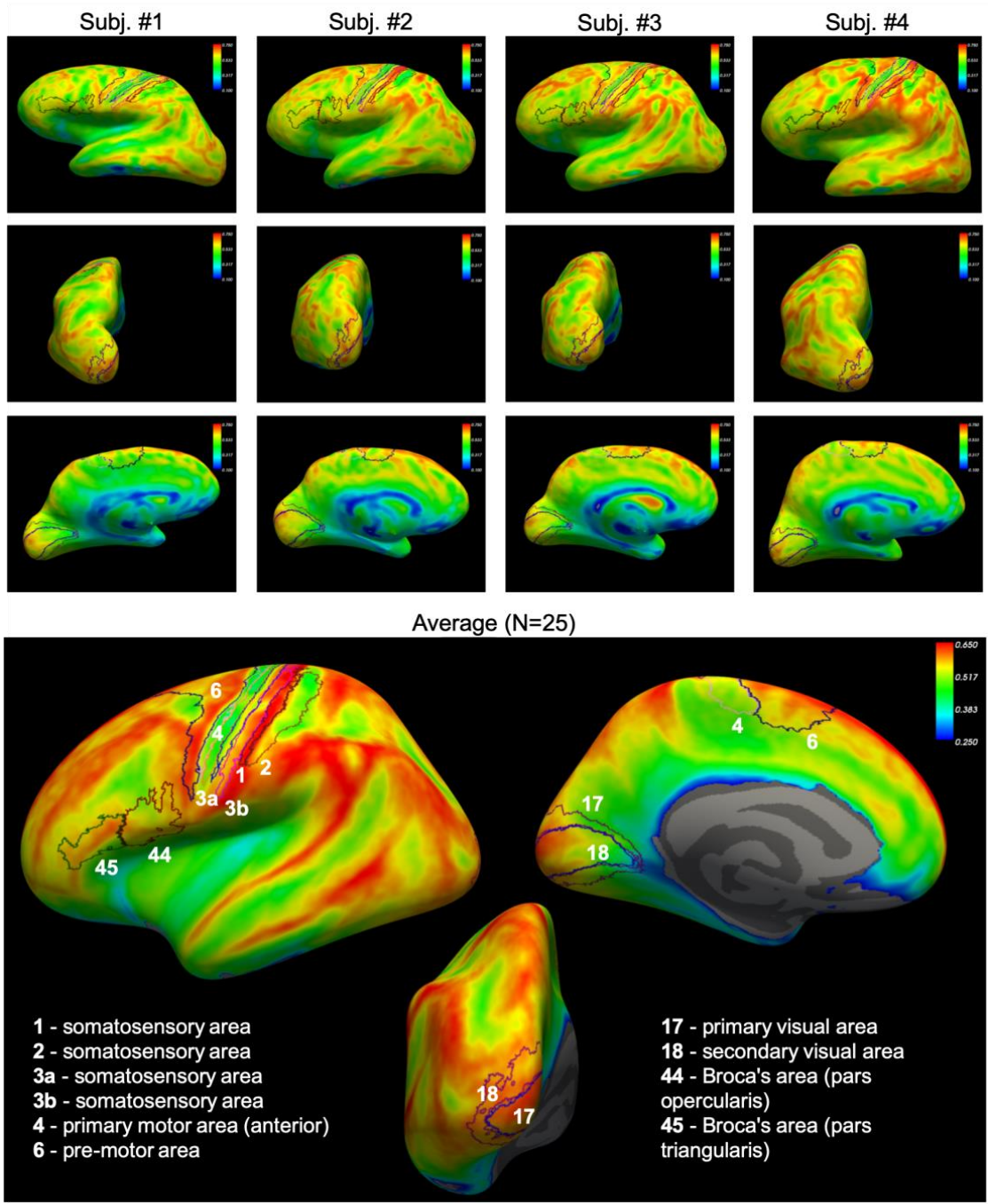

*Figure 11* Projection onto cortical surface (representing the whole cortical thickness) of quantitative maps of MR soma signal fraction $f_{is}$ for 4 representative subjects and for the average over the whole cohort of 25 healthy subjects. The principal Brodmann's areas available on FreeSurfer are also reported. We notice a remarkable correspondence between the boundaries of Brodmann areas and the gradient in $f_{is}$ values. This is particularly evident in the average map, while in the maps of individual subjects we can also appreciate sensible inter-subject variability.

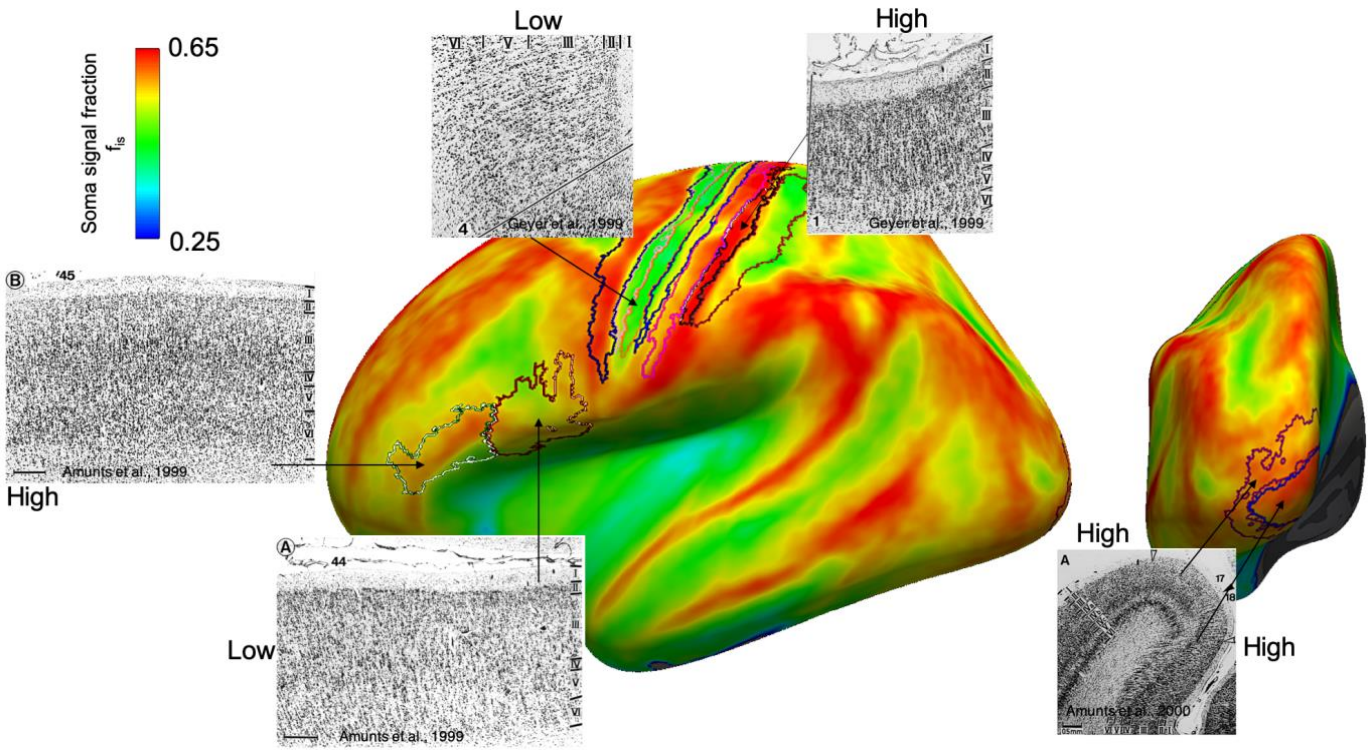

*Figure 12* *Projection onto cortical surface (representing the whole cortical thickness) of the quantitative map of MR soma signal fraction $f_{is}$ for the average over the whole cohort of 25 healthy subjects. In correspondence of the principal Brodmann areas we report histological images from literature, showing the typical soma arrangement and density used as criteria to delineate Brodmann areas. We find a very good correspondence between $f_{is}$ values and the expected pattern of soma density from histology (high, intermediate and low soma density as indicated).*

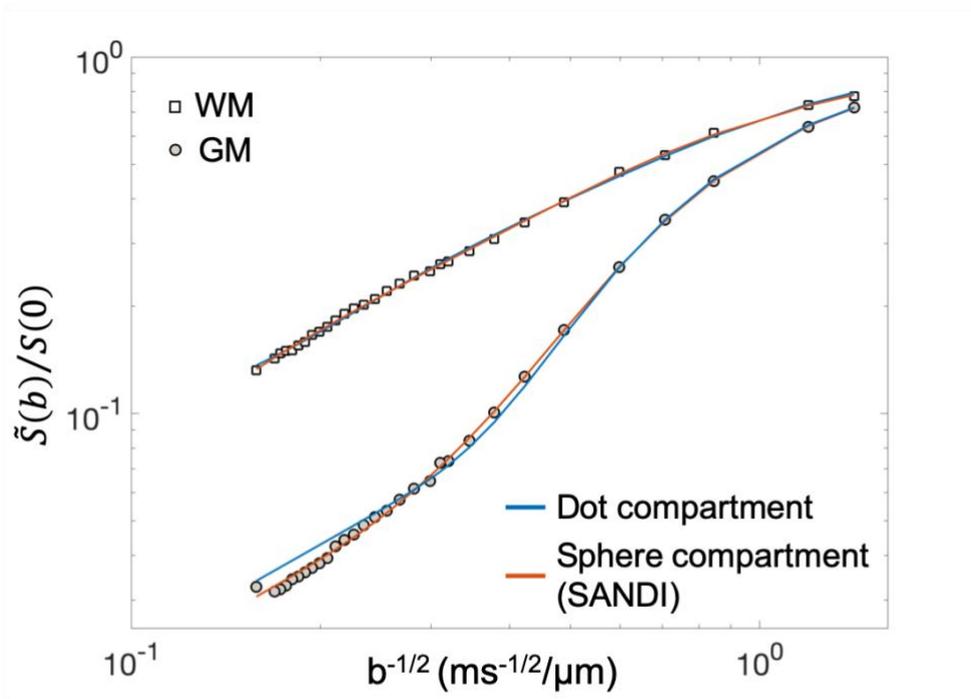

| | Dot compartment | | | | Sphere compartment (SANDI) | | | | |
|---|---|---|---|---|---|---|---|---|---|
| | $D_{in}$ (µm²/ms) | $f_{dot}$ | $f_{ec}$ | $D_{ec}$ (µm²/ms) | $D_{in}$ (µm²/ms) | $f_{is}$ | $r_s$ (µm) | $f_{ec}$ | $D_{ec}$ (µm²/ms) |
| **GM** | 2.97 | 0.00 | 0.58 | 0.54 | 1.26 | 0.61 | 5.5 | 0.35 | 1.26 |
| AICc | | 24.6 | | | | 16.5 | | | |
| **WM** | 1.85 | 0.05 | 0.12 | 0.20 | 1.65 | 0.05 | 2.3 | 0.12 | 0.31 |
| AICc | | 114.6 | | | | 54.4 | | | |

*Figure 13 Comparison of the SANDI model and the dot-compartment variant to describe measured DW-MRI signal decay at high b values. Two microstructure models were fitted to the data in **Figure 6**: the Eq. (10) from the SANDI model, and the Eq.(12) from the dot-compartment variant. The results of the fit for the DW-MRI data from the WM and GM ROIs in **Figure 6** are reported in the table, together with the values of the AICc (lower AICc indicates better fit).*

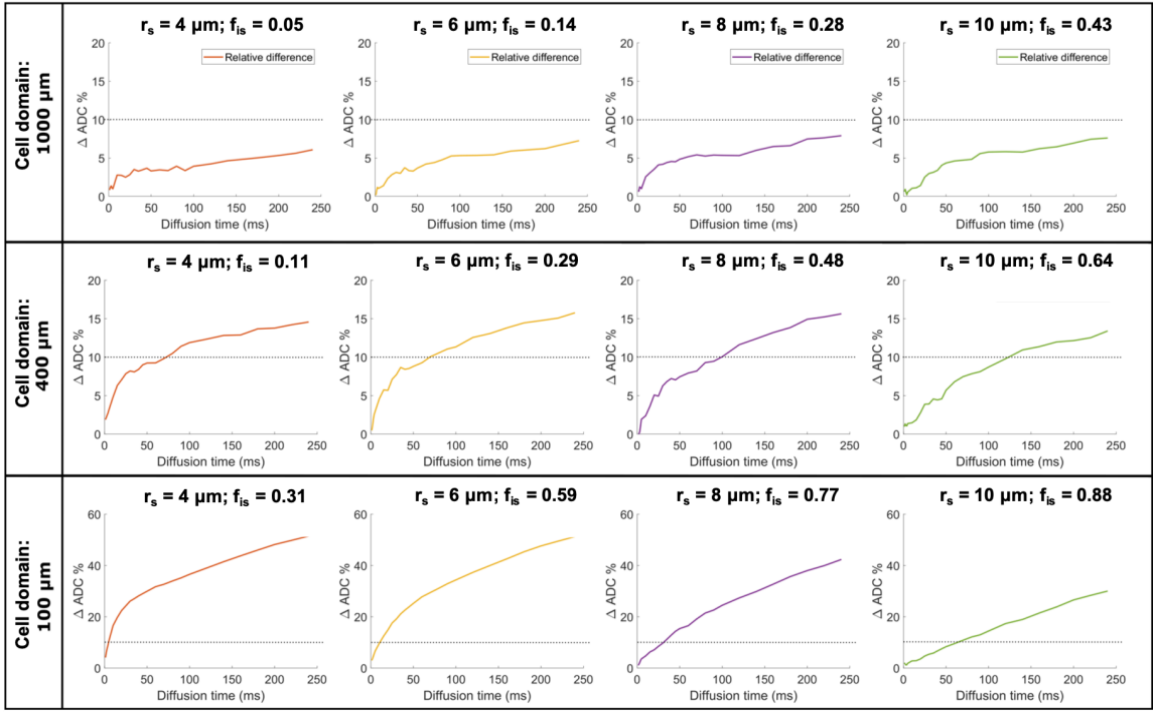

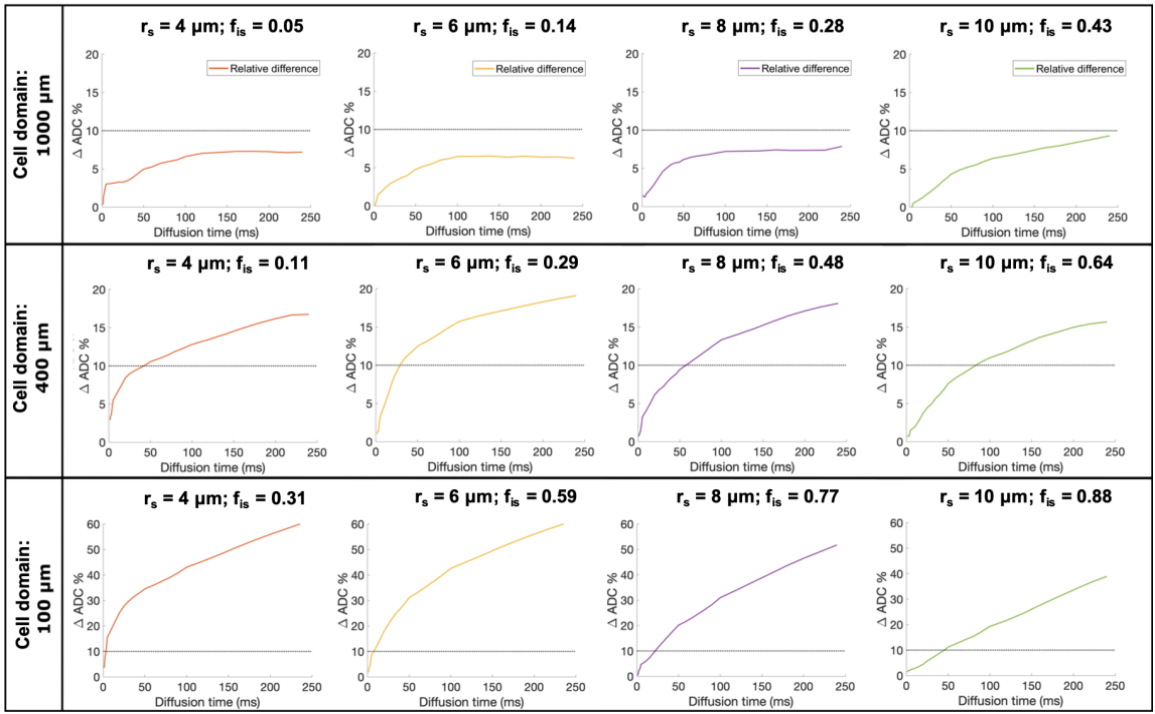

*Figure S1. Regime of validity of the compartment model for different bulk diffusivities. Relative percentage difference between the ADC in the exchange and non-exchange cases with bulk diffusivity $D_0 = 2$ $\mu m^2/ms$ (like in **Figure 3**) and $D_0 = 3$ $\mu m^2/ms$. The dashed lines show the 10% threshold used to define the diffusion time regime where the compartment model (10) is a reasonable approximation of cellular structures.*

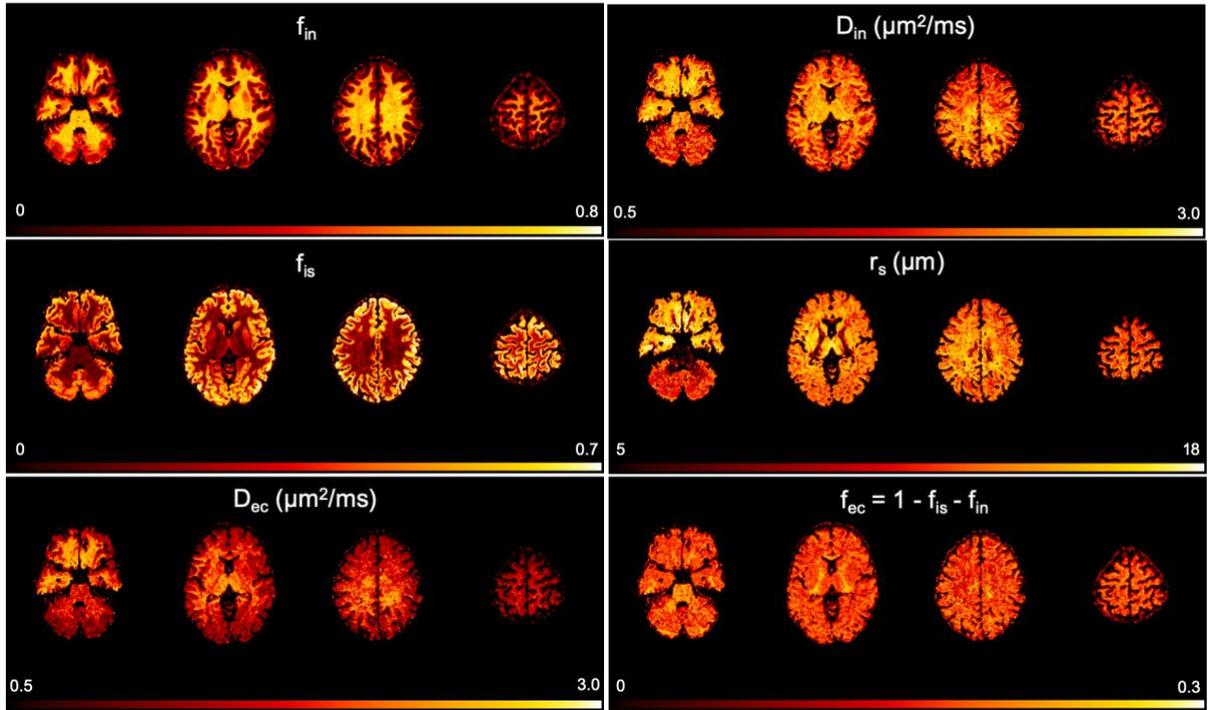

*Figure S2*. Example of SANDI parametric maps obtained by using a different training set from the one chosen in the Methods, section 3.3. In this case, the random forest regressor was trained using the following intervals for the model parameters: $f_{in}$=[0.01, 0.99]; $f_{ec}$=[0.01, 0.99]; $D_{in}$=[0.1, 3] $\mu m^2/ms$; $D_{ec}$=[0.1, 3] $\mu m^2/ms$; $r_s$=[1, 20] $\mu m$. Given the specific experimental protocol used in this study ($t_d$~20 ms), the measured DW-MRI signal is sensitive to soma size $r_s \leq 12$ μm. Therefore, training with $r_s$=[1, 20] μm is incorrect and leads to lower accuracy and precision in the model parameters prediction, especially for $r_s$, $D_{in}$ and $D_{ec}$, as evident from the comparison with **Figure 9**. In fact, we would expect lower $r_s$ in WM than in GM regions, and more uniform $D_{in}$ and $D_{ec}$ within WM and GM regions. While this is the case in **Figure 9**, it is clearly not the case here.